\documentclass[journal=dce]{Template/CUP-JNL-DTM}%

\usepackage{graphicx}
\usepackage{multicol,multirow}
\usepackage{amsmath,amssymb,amsfonts}
\usepackage{mathrsfs}
\usepackage{amsthm}
\usepackage{rotating}
\usepackage{appendix}
\usepackage{ifpdf}
\usepackage[T1]{fontenc}
\usepackage{newtxtext}
\usepackage{newtxmath}
\usepackage{textcomp}
\usepackage{xcolor}
\usepackage{lipsum}
\usepackage[colorlinks,allcolors=blue]{hyperref}
\usepackage{subcaption}

\theoremstyle{definition}

\numberwithin{equation}{section}

\jname{Data-Centric Engineering}
\articletype{ARTICLE}
\jyear{2025}

\begin{document}

\begin{Frontmatter}

\title[Article Title]{HFNO: an interpretable data-driven decomposition strategy for turbulent flows}

\author[1]{M. Cayuela}
\author[2]{P. Schmid}
\author[3]{V. Le Chenadec}
\author[4]{T. Sayadi}

\authormark{M. Cayuela \textit{et al}.}

\address[1]{\orgdiv{IJLRA}, \orgname{Sorbonne Université}, \orgaddress{\city{Paris}, \country{France}}} 

\address[2]{\orgdiv{Dept. Mechanical
Engineering}, \orgname{KAUST}, \orgaddress{\city{Thuwal}, \country{Saudi Arabia}}}

\address[3]{\orgdiv{MSME}, \orgname{Université Gustave Eiffel}, \orgaddress{\city{Champs-sur-Marne}, \country{France}}}

\address[4]{\orgdiv{M2N}, \orgname{Conservatoire National des Arts et Métiers}, \orgaddress{\city{Paris}, \country{France}}}

\authormark{M. Cayuela et al.}


	
\abstract{Fourier Neural Operators (FNOs) have demonstrated exceptional accuracy in mapping functional spaces by leveraging Fourier transforms to establish a connection with underlying physical principles. However, their opaque inner workings often constitute an obstacle to physical interpretability. This work introduces Hierarchical Fourier Neural Operators (HFNOs), a novel FNO-based architecture tailored for reduced-order modeling of turbulent fluid flows, designed to enhance interpretability by explicitly separating fluid behavior across scales. The proposed architecture processes wavenumber bins in parallel, enabling the approximation of dispersion relations and non-linear interactions. Inputs are lifted to a higher-dimensional space, Fourier-transformed, and partitioned into wavenumber bins. Each bin is processed by a Fully Connected Neural Network (FCNN), with outputs subsequently padded, summed, and inverse-transformed back into physical space. A final transformation refines the output in physical space as a correction model, by means of one of the following architectures: Convolutional Neural Network (CNN) and Echo State Network (ESN). We evaluate the proposed model on a series of increasingly complex dynamical systems: first on the one-dimensional Kuramoto–Sivashinsky equation, then on the two-dimensional Kolmogorov flow, and finally on the prediction of wall shear stress in turbulent channel flow, given the near-wall velocity field. In all test cases, the model demonstrates its ability to decompose turbulent flows across various scales, opening up the possibility of increased interpretability and multiscale modeling of such flows.}

\end{Frontmatter}

\section*{Impact Statement}
This work introduces the Hierarchical Fourier Neural Operator (HFNO) architecture, which decomposes the flow field into multiple spectral scales that are learnt and processed independently. This hierarchical treatment of scales significantly reduces model complexity while maintaining predictive accuracy in turbulent flow simulations and providing interpretability at various scales. Compared to standard Fourier Neural Operators, the HFNO achieves comparable statistical performance with five times fewer parameters.


\section{Introduction}
\label{sec:introduction}

Modeling high-dimensional dynamical systems governed by partial differential equations (PDEs), such as incompressible fluid flows or reactive transport phenomena, remains a central challenge in scientific computing. These systems exhibit complex nonlinear dynamics and span a broad range of spatial and temporal scales. Accurately resolving all relevant scales requires extremely high resolutions in both space and time, resulting in numerical models that routinely involve billions of degrees of freedom. As a result, high-fidelity simulations — such as Direct Numerical Simulation (DNS; ~\cite{moin_direct_1998}) or Large Eddy Simulation (LES; ~\cite{piomelli_large-eddy_1999}) — become prohibitively expensive, particularly at high Reynolds numbers~\citep{pope_turbulent_2000}. Even with modern high-performance computing infrastructures, such simulations often require massive parallelization and remain mostly restricted to limited time horizons or simplified geometries.

Recent advances in neural operator learning have opened new avenues for data-driven surrogate modeling of PDEs. Traditional neural network–based approaches, such as Physics-Informed Neural Networks (PINNs; ~\cite{raissi_physics-informed_2019}), model the solution by taking spatial and temporal coordinates as input and enforcing the governing equations through the loss function. While effective for low-dimensional problems, these methods require retraining for each new geometry or discretization, which limits their scalability. More recently, neural operator frameworks have been introduced to learn mappings directly between infinite-dimensional function spaces rather than pointwise quantities. Among them, DeepONet~\citep{lu_deeponet_2021} represents the operator using two sub-networks — a branch network sampling the input function at sensor points and a trunk network depending on the output coordinates. Although DeepONet provides theoretical guarantees for universal operator approximation, its accuracy can be sensitive to the choice and number of sensor points, as well as to the sampling of the training functions.

In contrast, the Fourier Neural Operator (FNO; \cite{li_fourier_2021}) achieves mesh and resolution invariance by parameterizing integral kernels in the Fourier domain. This formulation enables the learned operator to generalize across different discretizations of the same underlying function space, making FNOs particularly attractive for scalable, resolution-independent surrogate modeling of complex dynamical systems \citep{choubineh_fourier_2023, serrano_infinity_2023, wang_prediction_2024}. The FNO framework has since inspired a wide family of variants, including the Wavelet Neural Operator (WNO; \cite{tripura_wavelet_2023}), which improves multiscale feature representation through localized wavelet bases, and the Graph Neural Operator (GNO; \cite{li_neural_2020}), which extends operator learning to irregular domains and unstructured meshes. Despite their strong learning capabilities, Fourier Neural Operators (FNOs) remain deep architectures that offer limited interpretability. Although they embed physical knowledge through the use of Fourier transforms, it is still difficult to assign a clear physical meaning to the learned parameters after training.

This limitation motivates the design of a new architecture, inspired by FNOs, that retains the advantages of spectral representations while promoting a more interpretable and scale-aware structure. Turbulent flows are inherently multiscale phenomena, characterized by a broad range of interacting spatial and temporal structures. Traditional modeling approaches in fluid mechanics, such as Reynolds-Averaged Navier--Stokes (RANS; \cite{pope_turbulent_2000, launder_numerical_1974}) and Large-Eddy Simulation (LES; \cite{sagaut_pierre_large_2006,pope_turbulent_2000, meneveau_scale-invariance_2000}), explicitly exploit this separation of scales to design different modeling strategies for small- and large-small dynamics -- for instance, resolving the energy-containing eddies while modeling the subgrid-scale dynamics. This principle of scale separation has long proven effective in promoting both interpretability and consistency across models operating at different resolutions. Inspired by this idea, the present work aims to design a neural architecture that can automatically identify and model these scales separately, bridging physical insight from turbulence theory with modern operator-learning techniques. 

To this end, this contribution introduces the Hierarchical Fourier Neural Operator (HFNO), a novel deep learning architecture inspired by FNO, designed both to learn mappings between infinite-dimensional function spaces and to decompose predictions across different scales.
Our model divides the Fourier domain into disjoint wavenumber bins, each of which is processed by a dedicated Fully Connected Neural Network (FCNN;  \cite{hornik_multilayer_1989}). This enables the model to learn scale-specific nonlinear transformations, leading to improved generalization and physical interpretability. A final correction model -- implemented as either a CNN \citep{lecun_gradient-based_1998, ian_goodfellow_deep_2016} or an Echo State Network \citep{jaeger_echo_2010} -- further refines the output via a residual modeling.


The paper is organized as follows. Section~\ref{sec:neuraloperator} presents the general framework of operator learning, with a focus on FNOs. Section~\ref{sec:architecture} introduces the HFNO architecture proposed in this study. Section~\ref{sec:results} presents the results of our architecture compared to baseline methods when applied to several cases of variable complexity. Finally, conclusions of the work are discussed in Section~\ref{sec:conclusion}.

\section{Neural Operator}
\label{sec:neuraloperator}

In the general case of Operator learning, the goal is to learn a mapping between two infinite-dimensional function spaces. Let \( \mathcal{D} \subset \mathbb{R}^d \) be a bounded open set, and let \( \mathcal{U} = \mathcal{F}(\mathcal{D} ; \mathbb{R}^{d_u}) \) and \( \mathcal{Y} = \mathcal{F}(\mathcal{D} ; \mathbb{R}^{d_y}) \) be Banach spaces of functions taking values in \( \mathbb{R}^{d_u} \) and \( \mathbb{R}^{d_y} \) respectively. Here, $\mathcal{D} $ denotes the subset of the domain where the function is observed, and $\mathcal{U}$ and $\mathcal{Y}$ represent the input and output function spaces, respectively. The objective is therefore to learn a non-linear mapping $\mathcal{M} : \mathcal{U} \rightarrow \mathcal{Y}$ which may represent, for example, the solution of a PDE or the evolution of a trajectory, that we usually call an operator. 

To this end, we introduce Neural Operators, which are parametric numerical models denoted by $\mathcal{M}_{\theta}: \mathcal{U} \rightarrow \mathcal{Y}$. Suppose we are given a dataset of observation pairs $(u_i, y_i)_{i=1}^N \in (\mathcal{U} \times \mathcal{Y})^N$, such that $y_i = \mathcal{M}(u_i)$. The goal is to find parameters $\theta$ such that $\mathcal{M}_\theta$ approximates the true operator $\mathcal{M}$ as accurately as possible, in the sense of minimizing a cost function $J: \mathcal{Y} \times \mathcal{Y} \rightarrow \mathbb{R}_+$. This defines a supervised learning problem. 

In the present case, the function spaces involved are continuous. To enable numerical treatment, the domain $\mathcal{D}$ is discretized using a Cartesian mesh. Functions $u \in \mathcal{U}$ or $y \in \mathcal{Y}$ are then represented by their evaluations at a set of points $\mathcal{G} = (\mathbf{x}_j)_{j=1}^n \subset \mathcal{D} $, which sample the domain $\mathcal{D}$. For a Cartesian mesh, we typically define $n_k$ as the number of discretization points along each dimension $k \in \left \llbracket 1, d \right \rrbracket$, so that the total number of grid points is given by $n = \prod_k n_k$. The coordinates of the grid points in each dimension are specified by vectors $x^{(k)} \in \mathbb{R}^{n_k}$. Thus, any grid point can be identified by its multi-index $\left ( i _ 1, \dots, i _ d \right ) \in \left \llbracket 1,n_1 \right \rrbracket \times \cdots \times \left \llbracket 1, n_d \right \rrbracket$. 

In most cases, the learning process depends on the chosen discretization, and the evaluation of solutions is restricted to the discrete points $\mathbf{x}_j$, as is the case with DeepONets \citep{lu_deeponet_2021}. These points are commonly referred to as sensors. As we will see later, the use of the Fourier transform helps mitigate this dependence on specific sensor locations, enabling the design of mesh-invariant solvers.


One architecture that has shown remarkable performance and impact in recent years is the Fourier Neural Operator (FNO,~\cite{li_fourier_2021}). 
This architecture is defined iteratively through a sequence of mappings 
$v_0 \mapsto v_1 \mapsto \dots \mapsto v_T$, 
where $v_j \in \mathcal{F}(\mathcal{D}; \mathbb{R}^{d_v})$. 
The input field $u$ is first lifted to a higher-dimensional representation by a local transformation $P$, such that 
$\forall x, \ v_0(x) = P(u(x))$. 
The lifting operator $P$ is typically implemented as a shallow fully connected neural network (FCNN). Each transformation $v_t \mapsto v_{t+1}$, referred to as a Fourier layer, is defined as:
\begin{equation}
v_{t+1}(x) = \sigma \big( W_t v_t(x) + K_t(v_t)(x) \big),
\end{equation}
where $\sigma$ is a nonlinear activation function (typically a $\tanh$ or ReLU), 
$W_t \in \mathbb{R}^{d_v \times d_v}$ is a local linear transformation, 
and $K_t$ is an integral operator acting on the functional space $\mathcal{F}(\mathcal{D}; \mathbb{R}^{d_v})$, defined by
\begin{equation}
K_t(v_t)(x) = \mathscr{F}^{-1} \big( R_t \, \mathscr(v_t) \big)(x),
\end{equation}
where $\mathscr{F}$ and $\mathscr{F}^{-1}$ denote the Fourier and inverse Fourier transforms, respectively. 
The tensor $R_t$ parameterizes the operator $K_t$ and is defined by $R_t(k) \in \mathbb{C}^{d_v \times d_v}$. 
To better understand Equation (2), we can write explicitly 
\[
[R_t \mathscr{F}(v_t)](k) = R_t(k)\,\mathscr{F}(v_t)(k),
\]
which shows that the operation consists of a channel-wise multiplication in Fourier space. Finally, the output of the last layer $v_T$ is projected onto the output space through another shallow FCNN $Q$, yielding the final output $y(x) = Q(v_T(x))$. The reader is referred to the original paper \citep{li_fourier_2021} for more information regarding the details of the procedure.

As mentioned previously, a key advantage of the FNO architecture lies in its mesh-invariance: the operator acts globally over the entire spatial domain rather than locally on discrete points. All operations are either performed in physical space—independent of the specific spatial coordinates—or in Fourier space, where the low-frequency Fourier coefficients remain consistent across discretizations that satisfy the Nyquist–Shannon sampling criterion.



\section{Hierarchical Fourier Neural Operator}
\label{sec:architecture}

The Hierarchical Fourier Neural Operator (HFNO) operates on a latent space that differs from that of a classical autoencoder or a variational autoencoder \citep{hinton_reducing_2006, kingma_auto-encoding_2022}; instead, the latent space is defined in the Fourier domain to exploit physical structure, particularly in periodic problems. This choice not only aligns with the underlying structure of periodic problems, but also guarantees exact and lossless reconstruction of signals via analytically defined Fourier transform and inverse Fourier transform operations. In addition, the architecture treats the scales separately by performing a different transformation for different ranges of Fourier coefficients. The aim is to learn how to model phenomena specific to each range of scales, and to reduce the complexity of each component's task. 

Beyond its spectral formulation, the HFNO remains intrinsically meshless, operating independently of spatial discretization and thus adaptable to complex or irregular geometries. Although its design naturally aligns with periodic problems, several extensions have demonstrated its ability to handle non-periodic boundary conditions effectively \citep{li_fourier_2021, pathak_fourcastnet_2022, kovachki_neural_2024}.

\begin{figure}
    \centering
    \includegraphics[width=0.9\linewidth]{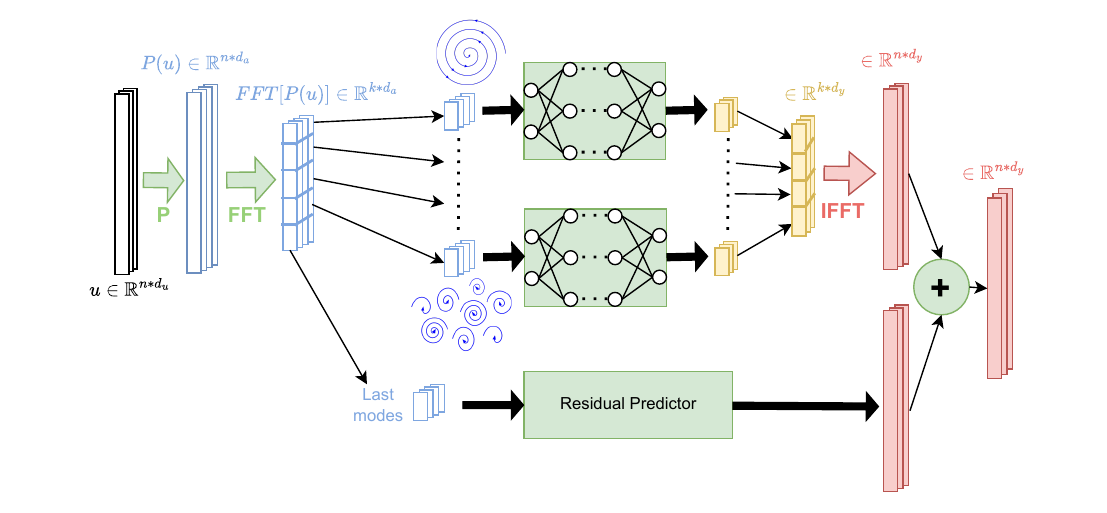}
    \caption{Generic architecture of HFNO proposed in this work.}
    \label{fig:architecture_H-FNO}
\end{figure}

The architecture proposed in this work is divided into two components: (i) the spectral block, which processes low-frequency modes grouped into blocks of submodes, and predicts coarse-scale behaviour, and (ii) the residual block, which processes the remaining modes in order to predict a correction term, i.e. residual closure. The second component will be discussed in terms of different design choices. A visualization of the generic HFNO architecture is shown in Figure~\ref{fig:architecture_H-FNO}. Each component will be further described in the following.

\subsection{Spectral block}

As in the case of the FNO, we first apply a lifting layer $P$ that projects the input $u$ into a higher-dimensional space, yielding $a \in \mathcal{F}(\mathcal{D}; \mathbb{R}^{d_a})$ with $d_a > d_u$, such that $\forall x \in \mathcal{D}, \, a(x) = P(u(x))$. The lifting operator $P$ is typically implemented as a shallow Fully Connected Neural Network (FCNN) without non-linear activation which amounts to a linear transformation from $\mathbb{R}^{d_u}$ to $\mathbb{R}^{d_a}$.

\begin{figure}
    \centering
    \includegraphics[width=0.9\linewidth]{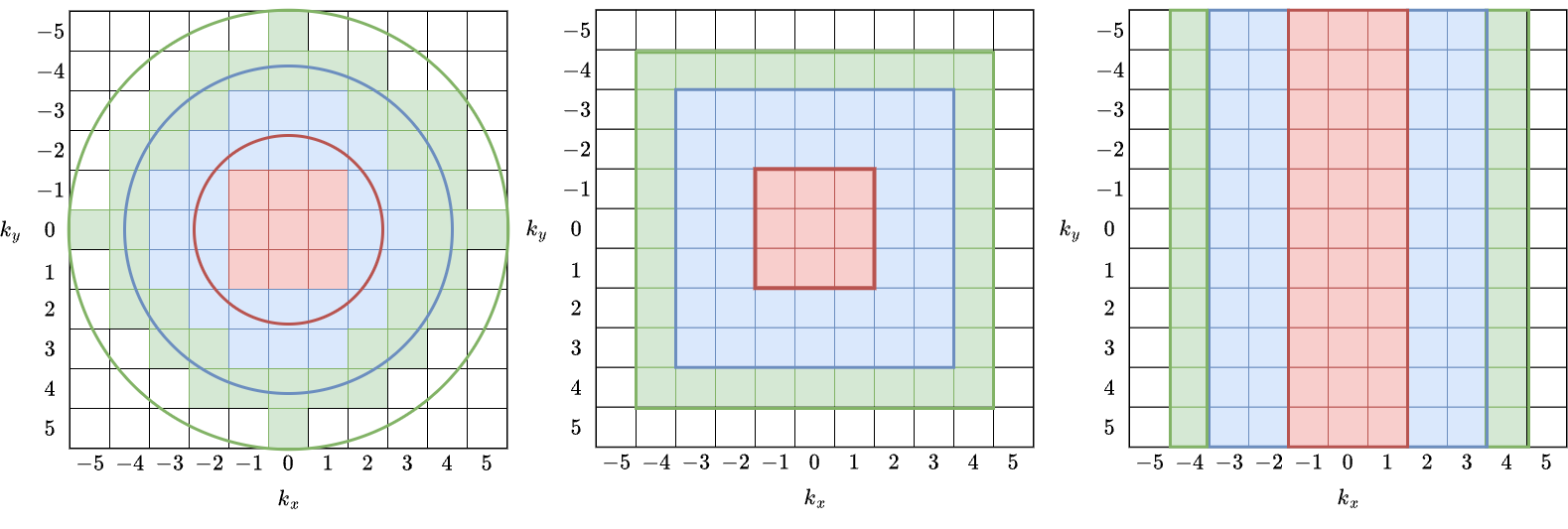}
    \caption{Three different partitioning strategies are shown from left to right: based on the $\ell_2$-norm, the maximum norm, and a semi-norm depending on a single axis.
}
    \label{fig:fourier_partitioning}
\end{figure}

The function $x \mapsto a(x)$ is then transformed into the frequency domain using a Fast Fourier Transform (FFT) over the spatial variable $x \in \mathcal{D}$. Using the previously introduced Cartesian grid notation, we numerically obtain $\hat{a}$, the Fourier representation of $a$, defined by $\mathbf{k} \mapsto \hat{a}(\mathbf{k}) \in \mathbb{C}^{d_a}$ for $\mathbf{k} \in \mathbb{Z}^d$. The frequency modes $\mathbf{k}$ are then partitioned into blocks. We select a sequence of positive real numbers $0 = c_0 < c_1 < \dots < c_p$, and define the $l$-th block of modes by
\[
\mathcal{K}_l = \left\{ \mathbf{k} \in \mathbb{Z}^d \,;\, c_{l-1} \leq \|\mathbf{k}\| < c_l \right\}, \quad \text{for } l \in \{1, \dots, p\}.
\]
The choice of the norm $\|\cdot\|$ is arbitrary, but the $\ell_2$-norm is typically used in practice. Examples of partitioning strategies are shown in Figure~\ref{fig:fourier_partitioning}.
Since we use an increasing sequence of thresholds to define the blocks, each block corresponds to a range of modes associated with progressively finer scales. This construction enables a flexible separation of scales, where the thresholds $c_l$ act as hyperparameters of the model.

The Fourier-transformed function $\hat{a}$ can now be divided into bins of Fourier modes according to the previously defined partitioning of Fourier space. These are defined as $\mathcal{A}_l = \{\hat{a}(\mathbf{k}) \mid \mathbf{k} \in \mathcal{K}_l\}$ for $l \in \{1, \dots, p\}$. A non-linear transformation is then applied independently to each of these bins, within a space that preserves the same Fourier modes but maps them to a new function valued in the output space. Concretely, each bin is flattened into a $\mathbb{C}$-valued vector of size $s_l = \mathrm{card}(\mathcal{K}_l) \cdot d_a$, and is mapped to a $\mathbb{C}$-valued vector of size $q_l = \mathrm{card}(\mathcal{K}_l) \cdot d_y$ using a mapping $g_l: \mathbb{C}^{s_l} \rightarrow \mathbb{C}^{q_l}$. We choose these mappings to be fully connected neural networks (FCNNs), for which we can configure hyperparameters such as depth and width. By allowing each mapping to be tuned independently, we can assign greater representational capacity to specific bins as needed. The output of each transformation now represents a bin of Fourier modes of a new function $\tilde{a}$, following the same partitioning used previously. We define this as $\tilde{\mathcal{A}}_l = \{\hat{\tilde{a}}(\mathbf{k}) \mid \mathbf{k} \in \mathcal{K}_l\}$. These bins are then aggregated using the inverse of the partitioning mapping, resulting in the complete Fourier representation of $\tilde{a}$. We finally inverse Fourier transform using an IFFT to go into the output space to obtain $\tilde{a}$ that will be the first part of the output, representing the lowest scales of the prediction. In the following, we refer to each scale-independent transform as a "Fourier layer".

As mentioned previously, this architecture is particularly well suited for periodic problems. However, it is possible to extend its applicability to non-periodic signals using the padding strategy. This involves applying zero-padding to the input signal, effectively enlarging the domain by adding zeros around the original data. By doing so, the signal can be treated as approximately periodic within the extended domain, enabling the model to exploit the Fourier-based latent space while mitigating boundary artifacts

\subsection{Residual Block}

The second part of the architecture serves as a residual corrector targeting the finest scales. We explore two different approaches. The first applies corrections directly in physical space, rather than in Fourier space as done in the previous stage. The second employs a Recurrent Neural Network (RNN) in Fourier space to incorporate memory and enhance the model’s ability to capture high-frequency components more effectively.

\subsubsection{Prediction in physical space}

The goal is to apply a transformation in physical space to account for the scales not handled by the first part of the model. To do so, we extract the Fourier modes that were not included in the previous partitioning, up to a user-defined threshold $c_{max}$. These modes define an additional block in the partition, given by $\mathcal{K}_{res} = \{\mathbf{k} \in \mathbb{Z}^d \mid c_p \leq \|\mathbf{k}\| < c_{max} \}$. We thus consider a truncated version of $\hat{a}$ defined as $\hat{a}_{res}: \mathcal{K}_{res} \rightarrow \mathbb{C}^{d_a}$. This representation is brought back to physical space via the inverse Fourier transform, yielding $a_{res}$ with values in $\mathbb{R}^{d_a}$, i.e., $a_{res}(\mathbf{x}) \in \mathbb{R}^{d_a}$. \\
The final step consists in mapping this latent-space representation to the output space. This transformation is of the form $\mathcal{W}: \mathcal{F}(\mathcal{D}; \mathbb{R}^{d_a}) \rightarrow \mathcal{F}(\mathcal{D}; \mathbb{R}^{d_y})$. We investigate two types of transformations. The first is a simple linear map $W: \mathbb{R}^{d_a} \rightarrow \mathbb{R}^{d_y}$, constant in space, such that $\forall x \in \mathcal{D}, \; W a_{res}(x) = \tilde{a}_{res}(x)$. Due to its limited expressiveness, we also explore a more flexible model based on a convolutional neural network (CNN), which operates directly on the full field $a_{res}$ rather than pointwise on $a_{res}(x)$.\\
Finally, the two contributions are summed to produce the final prediction: $\tilde{y} = \tilde{a} + \tilde{a}_{\mathrm{res}}$.

\subsubsection{Prediction in wave-number space}

Echo State Networks (ESNs,~\cite{jaeger_echo_2010}) are a specific class of Recurrent Neural Networks (RNNs) belonging to the reservoir computing paradigm. They aim to simplify the training of recurrent architectures while preserving their capability to model complex temporal dynamics. Unlike traditional RNNs, where all recurrent weights are trained, ESNs employ a large, randomly connected, and fixed recurrent layer known as the reservoir. Only the output connections are learned, resulting in efficient and numerically stable training. ESNs have been widely employed in the scientific community to predict dynamical trajectories, both in full and reduced-order spaces \citep{ozalp_stability_2025, heyder_echo_2021}. Their proven ability to capture complex temporal dependencies motivates their consideration for future extensions of the proposed model, particularly in the context of predicting dynamical trajectories.

\begin{figure}
    \centering
    \includegraphics[width=0.6\linewidth]{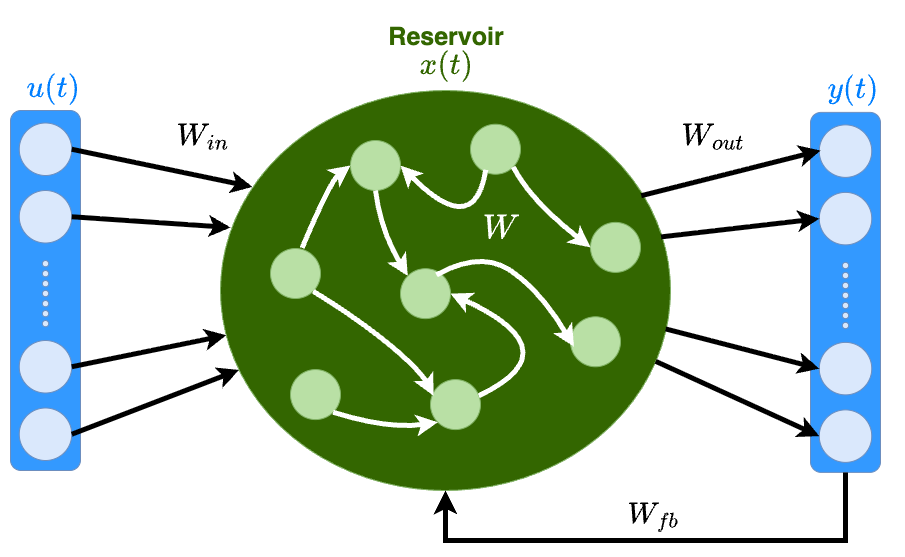}
    \caption{Architecture of an Echo State Network (ESN)}
    \label{fig:esn_architecture}
\end{figure}

An ESN is composed of three main components, illustrated in Figure~\ref{fig:esn_architecture}: an \textbf{input layer}, which projects the external input signal into the high-dimensional reservoir space; a \textbf{reservoir}, a sparsely connected recurrent network providing rich nonlinear dynamics; and an \textbf{output layer}, which linearly combines the reservoir states to generate the network output.

Let $\mathbf{u}(t) \in \mathbb{R}^{N_u}$ denote the input vector at time step $t$, 
$\mathbf{x}(t) \in \mathbb{R}^{N_x}$ the reservoir state vector, and 
$\mathbf{y}(t) \in \mathbb{R}^{N_y}$ the output vector. 
The ESN dynamics are governed by the following discrete-time equations:
\begin{align}
\tilde{\mathbf{x}}(t+1) &= f\!\left( W_{\mathrm{in}}\mathbf{u}(t+1) + W \mathbf{x}(t) + W_{\mathrm{fb}}\mathbf{y}(t) \right), \\
\mathbf{x}(t+1) &= (1-\alpha)\,\mathbf{x}(t) + \alpha\,\tilde{\mathbf{x}}(t+1), \\
\mathbf{y}(t+1) &= W_{\mathrm{out}}
\begin{bmatrix}
\mathbf{x}(t+1) \\[2pt]
\mathbf{u}(t+1) \\[2pt]
1
\end{bmatrix},
\end{align}
where:
\begin{itemize}
    \item $W_{\mathrm{in}} \in \mathbb{R}^{N_x \times N_u}$ is the input weight matrix,
    \item $W \in \mathbb{R}^{N_x \times N_x}$ is the recurrent (reservoir) weight matrix,
    \item $W_{\mathrm{fb}} \in \mathbb{R}^{N_x \times N_y}$ is an optional feedback matrix,
    \item $W_{\mathrm{out}} \in \mathbb{R}^{N_y \times (N_x + N_u + 1)}$ is the trainable output matrix,
    \item $f(\cdot)$ is a nonlinear activation function, typically $\tanh$,
    \item $\alpha$ is the leaking rate.
\end{itemize}
During training, the internal weights ($W_{\mathrm{in}}$, $W$, $W_{\mathrm{fb}}$) remain fixed, while $W_{\mathrm{out}}$ is optimized using linear or ridge regression to minimize the prediction error. The matrices $W_{\mathrm{in}}$ and $W$ are typically initialized randomly, with $W_{\mathrm{in}}$ being dense and $W$ sparse.

The reservoir acts as a nonlinear dynamical system that maps the input sequence into a high-dimensional state space. Its key property, the \textit{echo state property} (ESP), ensures that the influence of past inputs on the current state vanishes over time. Formally, the ESP requires that, for any bounded input sequence $\{\mathbf{u}(t)\}$, the difference between two state trajectories $\mathbf{x}_1(t)$ and $\mathbf{x}_2(t)$, initialized differently but driven by the same input, converges to zero as $t \to \infty$. A sufficient (though not necessary) condition for the ESP is that the spectral radius $\rho(W)$ of the reservoir weight matrix satisfies:
\begin{equation}
\rho(W) < 1,
\end{equation}
which guarantees that the internal dynamics are contractive and stable. ESNs thus provide an efficient and robust framework for modeling temporal and sequential data. Their main advantages include fast training, numerical stability, and the ability to capture complex temporal dependencies through high-dimensional nonlinear reservoir dynamics. They have been successfully applied to time-series prediction, signal processing, system identification, and control tasks.

Echo State Networks (ESNs) are particularly well suited for low-dimensional data, meaning that the input and output spaces, where $u$ and $y$ lie, should have relatively small dimensions. This is because the expressive power of ESNs relies on the large size of the reservoir compared to the input and output spaces. Moreover, the trainable parameters are obtained through a least-squares optimization, whose computational complexity grows linearly with the output dimension. For this reason, ESNs cannot be applied directly in the physical space, which typically involves an extremely high number of degrees of freedom, often reaching billions in practical fluid dynamics problems. A reduced representation of the data is therefore required. Following the methodology introduced in the previous section, the Fourier transform of the physical-space data provides a suitable reduced representation. Consequently, we use the remaining high-frequency modes — those not already employed in the Fourier layers — as the input for the ESN residual predictor.

\subsection{Training process}

Regarding the different nature of the two architectures, we adopt distinct training procedures for each. In both cases, the cost function $\mathcal{J}$ is defined as the $L^2$ norm integrated over the domain $\mathcal{D}$. Given two fields $y$ and $\tilde{y}$, the loss is expressed as

\begin{equation}
    \mathcal{J}(y,\tilde{y}) = \frac{1}{\left \vert \mathcal{D} \right \vert} \|y - \tilde{y}\|_{L^2}^2 = \frac{1}{\left \vert \mathcal{D} \right \vert} \|y(\mathbf{x}) - \tilde{y}(\mathbf{x})\|^2 \, d\mathbf{x} \approx \frac{1}{n} \sum_{j=1}^n \|y(\mathbf{x}_j) - \tilde{y}(\mathbf{x}_j)\|^2,
\end{equation}
where $\|\cdot\|$ denotes the Euclidean norm in the codomain of $y$ and $\left \vert \mathcal{D} \right \vert$ the volume of $\mathcal{D}$.

Due to the limited number of trainable parameters in the ESN and the linearity of its output layer, training can be efficiently performed using a least squares approach, which is both faster and more accurate than the gradient-based methods commonly employed in standard deep learning models. Consequently, the residual component is trained separately using the ESN. In cases where the ESN is not used, all model components are trained jointly in an end-to-end manner.

In cases where the residual corrector is either linear or implemented as a CNN, we are given a dataset of input/output pairs ${(u_j, y_j)}_{j=1}^{N_{data}}$ and aim to approximate the operator $\mathcal{M}$ mapping $u$ to $y$ using a parameterized model $\mathcal{M}_\theta$, constructed as previously described. The objective is to minimize the loss $\sum_j \mathcal{J}(y_j, \mathcal{M}_\theta(u_j))$ over the parameter space of $\theta$. The model was trained using the Adam optimizer, which belongs to the family of stochastic gradient-based optimization methods.

In the case where the residual corrector is an ESN, the training procedure differs slightly. Let $\mathcal{M}^{spec}$ denote the spectral binned component of the model, and $\mathcal{M}^{res}$ the residual correction component. We first train $\mathcal{M}^{spec}$ by minimizing the cost function. Once trained, we use it to compute predictions $y_{pred,j} = \mathcal{M}^{spec}(u_j)$. The ESN-based corrector $\mathcal{M}^{res}$ is then trained to predict the residual $y_j - y_{pred,j}$ using the ESN training procedure.

\section{Results}
\label{sec:results}
The results are presented for three distinct cases, each selected to reflect varying levels of complexity and application objectives. The first case involves the Kuramoto–Sivashinsky (KS) equation, providing a simplified scenario that facilitates a baseline comparison of the proposed method. In the second case, involving the Kolmogorov flow, the model is employed as an emulator to predict subsequent snapshots in a time-resolved sequence, thereby demonstrating the suitability of Echo State Networks (ESNs) for temporal forecasting tasks. Finally, in the turbulent channel flow case, the model is utilized to map filtered data from an off-wall plane to wall shear stress, serving as an a priori test for wall model implementation in large-eddy simulations.

\subsection{Kuramoto-Sivashinsky equation}

We first evaluate our model on the one-dimensional Kuramoto–Sivashinsky (KS) equation, a standard benchmark in nonlinear dynamics due to its ease of numerical simulation and its rich nonlinear behavior. The KS equation is formulated as follows:
\begin{equation}
\left\{
\begin{aligned}
    &\frac{\partial u}{\partial t} = -\frac{1}{2} \nabla \cdot u^2 - \Delta u - \nu \Delta^2 u, \\
    &u(x+L,t) = u(x,t), \\
    &u(x,0) = g(x),
\end{aligned}
\right.
\end{equation}
where $L$ denotes the length of the one-dimensional domain, $\nu$ controls the dissipative properties of the system, and $g$ represents the prescribed initial condition. The first equation can be reorganized as
\[
\frac{D u}{D t} = -\Delta u - \nu \Delta^2 u,
\]
where
\[
\frac{D u}{D t} = \frac{\partial u}{\partial t} + \frac{1}{2} \nabla \cdot u^2
\]
is the material derivative. Our objective is to predict the material derivative as a function of $u$. In particular, we aim to learn the operator
\[
u \mapsto -\Delta u - \nu \Delta^2 u,
\]
which is a linear operator parameterized by $\nu$.

To generate the datasets, two numerical simulations of the Kuramoto--Sivashinsky (KS) equation are performed using different initial conditions. The initial conditions are periodic signals that are represented as weighted sums of sine and cosine functions. Different simulations are generated by varying these coefficients. The spatial domain is defined by $L = 22$, and the viscosity parameter is set to $\nu = 0.7$. Each simulation produces a trajectory consisting of $N_t = 2000$ temporal snapshots on a spatial grid with $n_x = 1024$ points. The first trajectory is used for training and the second for testing.

We employ an HFNO architecture composed of two parallel Fourier layers and a linear residual predictor. In this configuration, the first layer receives the first six Fourier modes, while the second layer processes the subsequent six modes. The choice of the number of layers and the allocation of modes was determined by comparing different hyperparameter configurations. As expected, increasing the number of layers led to a decrease in accuracy, due to the loss of interconnections between modes and the resulting disruption of dependencies within the flow. Therefore, we selected a two-layer configuration, which yielded the minimal error on the training set. Each layer is implemented as an MLP consisting of two dense layers with 64 neurons, using GeLU as the activation function. The lifting dimension is set to 64. 

\begin{figure}
    \centering
    \includegraphics[width=0.95\linewidth]{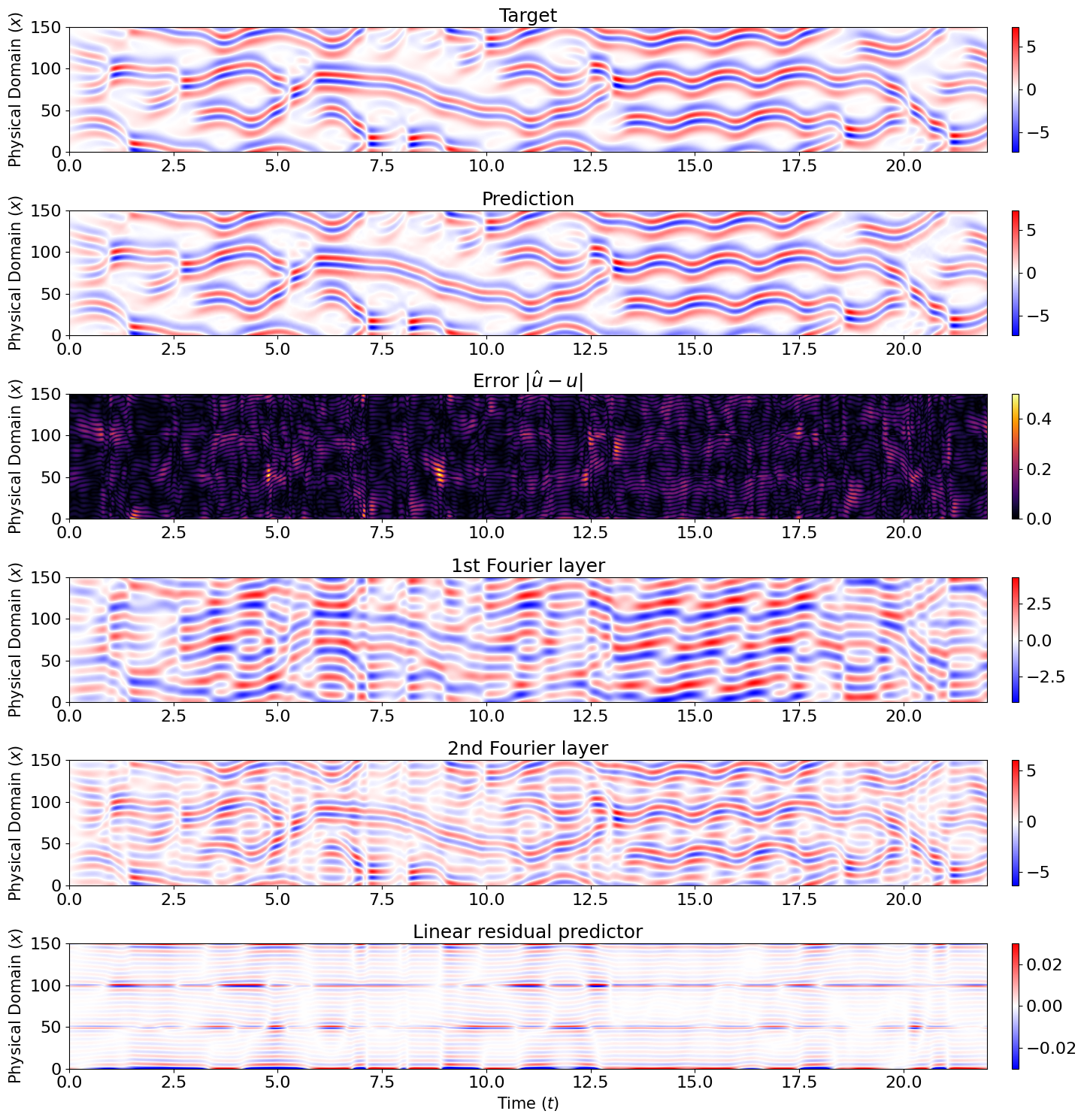}
    \caption{From top to bottom: target flow, prediction, absolute value of the error, contributions of each Fourier layer and residual predictor}
    \label{fig:ks_eq_preds}
\end{figure}

\begin{figure}
    \centering
    \includegraphics[width=0.6\linewidth]{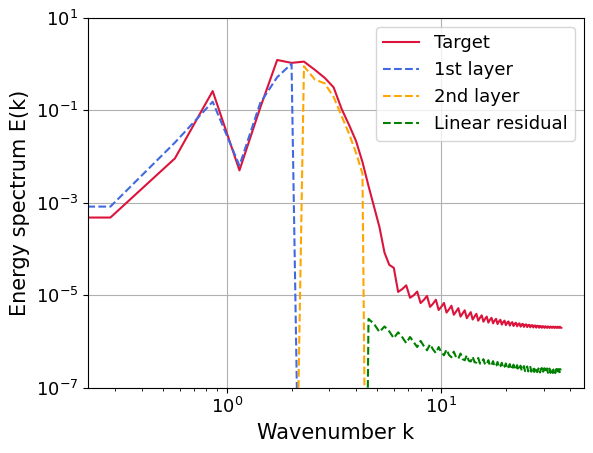}
    \caption{Kuramoto-Sivashinsky: energy spectrum of the raw data and the prediction of each part of the model}
    \label{fig:energy_spectrum_KS}
\end{figure}

Figure~\ref{fig:ks_eq_preds} shows the model predictions on test data for $\nu=0.7$, compared against the ground truth. The corresponding error field is also reported. The last three plots are the contribution of each component of the architecture to the final prediction. As expected, the first Fourier layer predominantly captures the large-scale structures, while the second layer resolves the smaller-scale features, while the residual predictor accounts for the remaining discrepancies. To complement these results, Figure~\ref{fig:energy_spectrum_KS} presents the energy spectrum of the Kuramoto--Sivashinsky testing trajectory, along with the energy spectra of the predictions from each component of the model (the two Fourier layers and the linear residual predictor). As shown in the figure, each Fourier layer predominantly reconstructs the flow features within the spectral range it was designed to represent, thereby capturing the associated structures at those characteristic scales. This observation substantiates that the HFNO architecture achieves the intended scale separation: different layers operate on complementary regions of the energy spectrum, collectively enabling an accurate multi-scale representation of the chaotic flow dynamics.

\subsection{Kolmogorov flow}

In this section, we analyze the performance of our model on the Kolmogorov flow. We apply it to two distinct simulations conducted at different Reynolds numbers. The Kolmogorov flow describes the motion of an incompressible fluid in a two-dimensional periodic domain, driven by a sinusoidal forcing. It is governed by the incompressible Navier–Stokes equations:
\begin{equation}
\left\{
\begin{aligned}
    &\nabla \cdot \mathbf{u} = 0, \\
    &\frac{\partial \mathbf{u}}{\partial t} + \mathbf{u} \cdot \nabla \mathbf{u} = -\nabla p + \nu \nabla^2 \mathbf{u} + \mathbf{f} 
\end{aligned}
\right.
\end{equation}
where $\mathbf{u}$ is the velocity field, $p$ is the pressure, $\nu$ is the kinematic viscosity, and $\mathbf{f}$ is the external forcing term. In the classical Kolmogorov setup, the forcing is unidirectional and sinusoidal in space:
\begin{equation}
    \mathbf{f}(x, y) = (F_0 \sin(kx), 0),
\end{equation}
where $F_0$ is the forcing amplitude and $k = \frac{2\pi}{L}$ is the wavenumber corresponding to the periodicity length $L$ of the domain in the $x$-direction. The domain is typically taken as a two-dimensional square periodic domain, $\Omega = [0, L]^2$, with periodic boundary conditions in both directions. This setup allows for the study of transition to turbulence, nonlinear energy transfer, and the behavior of turbulence models in a controlled setting. 

The overall objective of our model is to predict the velocity field at time $t+1$ given the configuration at time $t$. To do so, we generate simulations of Kolmogorov using a pseudospectral solver using a Fourier-Galerkin approach \citep{canuto_spectral_1988, magri_luca_github_2021}. The initialization is chosen randomly, and the simulation is run without saving snapshots during the transient phase. We then obtain a sequence of snapshots ${\mathbf{u}_0, \ldots, \mathbf{u}_N}$, where $\mathbf{u}_0$ corresponds to the state recorded at time $t_0 = 0$ (i.e., the end of the transient phase). For all $i$, $\mathbf{u}_i$ denotes the snapshot recorded at time $t_i = i\times\Delta t$. The total simulation time is therefore $T = N\times\Delta t$.

\begin{figure}
    \centering
    \includegraphics[width=0.8\linewidth]{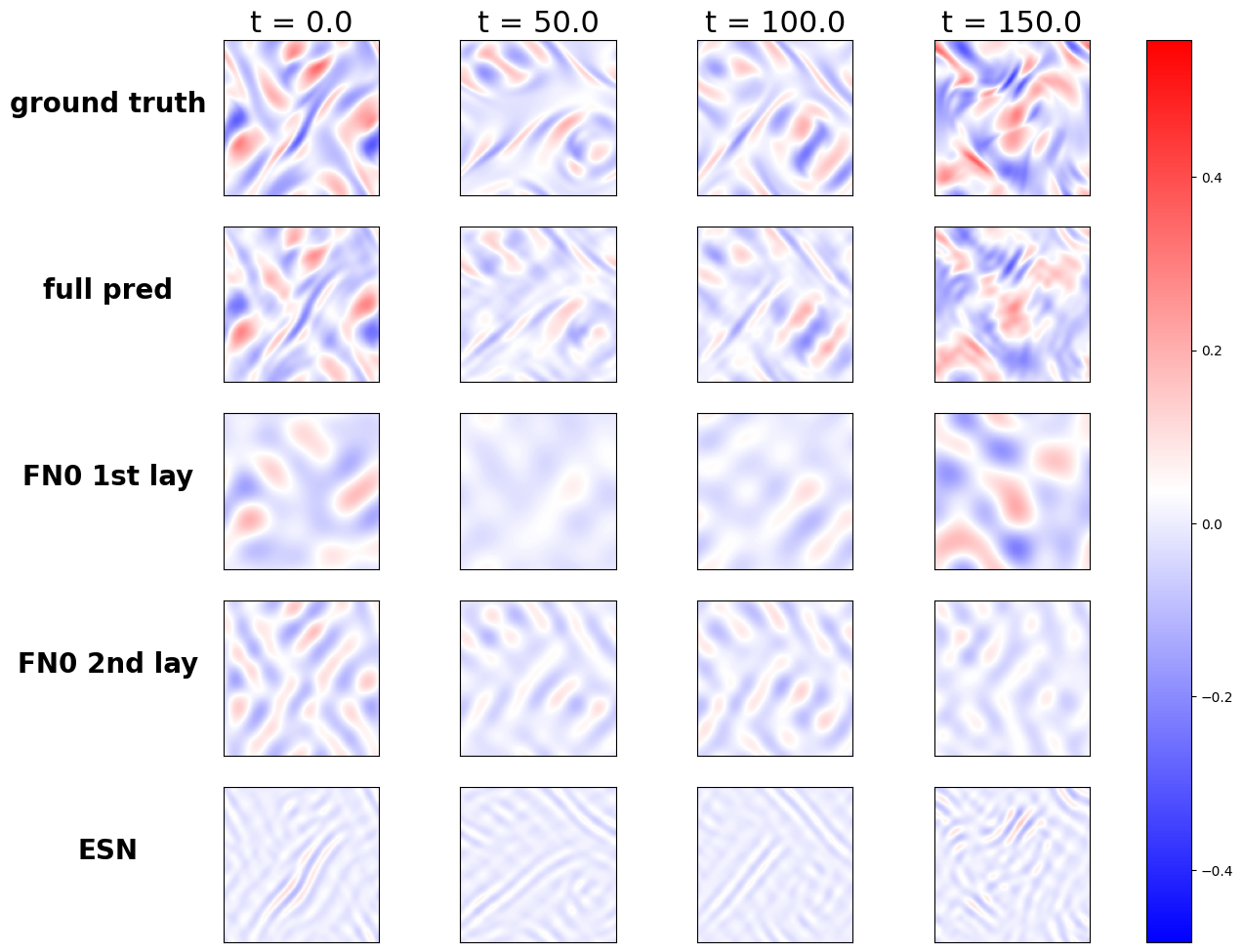}
    \caption{Vorticity predicted by each part of the model compared to the ground truth for different snapshots; Kolmogorov flow at $Re=90$.}
    \label{fig:RE90vorticity}
\end{figure}

In this case, we employ an HFNO architecture composed of two parallel Fourier layers. The scale separation is performed using the $L_2$ norm: modes $\mathbf{k}$ such that $||\mathbf{k}||_{L_2} \in [0,4]$ are assigned to the first layer, while those with $||\mathbf{k}||_{L_2} \in ]4,8]$ are assigned to the second. This separation follows the classical structure of turbulent flows: the first layer captures the energy-containing scales, corresponding to the largest coherent structures, while the second layer focuses on the inertial range, where energy is transferred nonlinearly between scales. By aligning the Fourier decomposition with these physically meaningful ranges, each layer is tasked with learning the dynamics specific to its scale, improving both interpretability and efficiency. The MLPs used for each layer are designed as in the KS equation experiment. Since the underlying system is dynamical, we adopt a variant of HFNO in which the residual predictor is implemented as an Echo State Network (ESN). We evaluate this architecture at Reynolds number $Re = 90$. While Although additional tests were conducted at a lower Reynolds number of 
$Re = 34$ (not shown), only the higher Reynolds number results are presented here, as they produce an energy spectrum exhibiting a more extensive cascade, thereby facilitating clearer identification of distinct flow scales.

Figure~\ref{fig:RE90vorticity} shows the vorticity fields computed from the predictions of each component of the model, namely the two Fourier layers and the ESN residual predictor. These predictions are visually compared with the ground truth at different snapshots. As in the previous case, the results highlight how the different layers separate the spatial scales.

\begin{figure}
    \centering
    \includegraphics[width=0.8\linewidth]{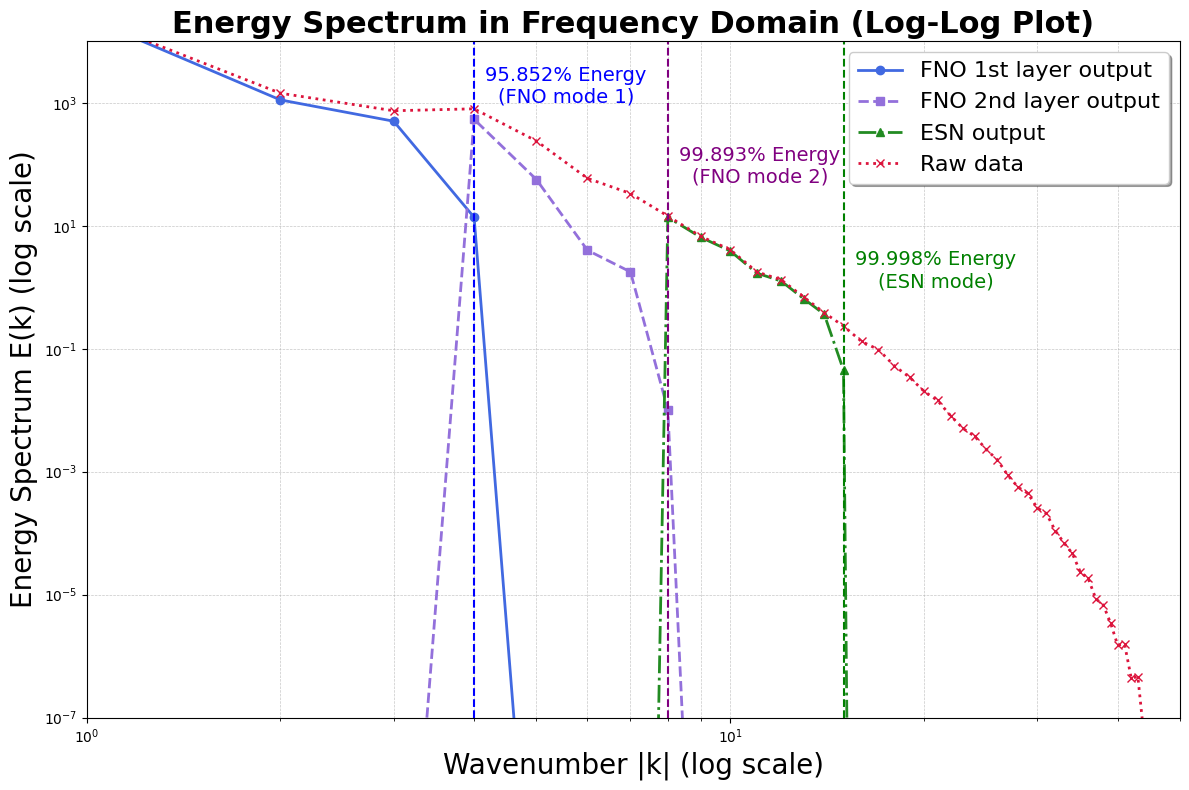}
    \caption{Energy spectrum of the raw data and the prediction of each part of the model; Kolmogorov flow at $Re=90$. }
    \label{fig:RE90energyspectrum}
\end{figure}

In addition, Figure~\ref{fig:RE90energyspectrum} presents the energy spectrum associated with each component of the model. As expected, each component primarily contributes to its corresponding range of scales: the first Fourier layer dominates the low-wavenumber energy-containing scales, the second layer captures the intermediate inertial-range dynamics, and the ESN residual accounts for smaller-scale structures and corrections, ensuring a faithful reconstruction across the full spectrum.

\subsection{Turbulent channel flow at $Re_{\tau} = 180$}

We next address a more challenging problem involving turbulent channel flow, with a specific focus on wall modeling. In this context, the objective is to infer accurate wall shear stresses from filtered flow data obtained at an off-wall plane, which can then serve as boundary conditions at the wall—thereby enabling coarser near-wall resolution in simulations. Direct Numerical Simulation (DNS) of wall-bounded turbulence becomes intractable at high Reynolds numbers, as the near-wall structures require extremely fine grids. Large-Eddy Simulation (LES) reduces the computational cost by resolving only the large scales, but the near-wall region remains too demanding to capture explicitly~\citep{choi_grid-point_2012}. Wall models provide an efficient remedy by replacing the no-slip condition with a closure that accounts for the unresolved part of the boundary layer~\citep{bae_dynamic_2019, cabot_approximate_2000}.
Using the HFNO  architecture, our goal is to learn the underlying operator that maps off-wall filtered data to the corresponding wall shear stress. The broader motivation is that, by successfully learning this operator, we can parameterize it and generalize its application to varying flow conditions. This approach represents a significant advancement beyond traditional data-driven models, offering a more general and physically-informed framework than simple input–output mappings.

To evaluate the performance of our proposed HFNO model, we first reproduce the experimental setup and results of \cite{dupuy_data-driven_2023}, ensuring that our implementation closely follows their architecture, training procedure, and data preprocessing. The reproduced results are consistent with those reported in the original work, providing a reliable baseline for comparison. We then compare our model's predictions to this baseline to assess improvements in accuracy and scale-specific interpretability in an a priori test condition. The baseline model selected for comparison is a Multilayer Perceptron (MLP), which predicts the wall shear stress $\tau$ locally. Specifically, it uses the velocity field sampled at a given distance from the wall and within a local neighborhood surrounding each target point to produce the corresponding $\tau$ prediction.

The data considered in this numerical experiment consists of snapshots of the DNS of a channel flow at a friction Reynolds number; $\operatorname{Re} _ {\tau} = 180$. The numerical simulation were performed using the SIMSON flow software, a pseudo-spectral solver for incompressible Navier-Stokes equations \citep{chevalier_mattias_simson_2007}.
The domain size is $8h\pi \times 2h \times 4h\pi$ and the geometry is periodic in streamwise and spanwise directions and the cartesian meshgrid gives snapshots with dimensionality $n_x \times n_y \times n_z$ where $n_x = 193$, $n_y = 65$ and $n_z = 193$.


For each snapshot, we compute the wall shear stress tensor defined as  
\begin{equation}
\overrightarrow{\mathbf{\tau}} = \Sigma \cdot e_n - \big(e_n \cdot \Sigma \cdot e_n\big) \, e_n ,
\end{equation}
where
\begin{equation}
\Sigma = \mu \left( \nabla u + (\nabla u)^T - \tfrac{2}{3} (\nabla \cdot u)\, I_d \right) .
\end{equation}
Our objective is to predict the norm of the wall shear stress,
\begin{equation}
\tau = \|\overrightarrow{\mathbf{\tau}}\| .
\end{equation}
The model is designed to predict $\tau$ using as input the velocity field evaluated at any distance $\Delta y$ from the wall. To this end, from a single snapshot we generate multiple velocity fields corresponding to different distances from the wall, restricted to the range $\Delta y^+ \in [25, 50]$, where the notation $(\cdot)^+$ refers to wall units. The conversion from physical units to wall units is given by
\begin{equation}
\Delta y^+ = \Delta y \, \sqrt{\tfrac{\tau}{\rho}} \, \tfrac{1}{\nu}.
\end{equation}
Since we do not wish to explicitly provide the model with the information about the distance $\Delta y$, we instead normalize the pairs $(u(\Delta y), \tau)$ with respect to $\Delta y$, such that
\begin{equation}
(\tilde{u}, \tilde{\tau}) = \left( \rho \left(\tfrac{\Delta y}{\mu}\right) u(\Delta y), \; \rho \left(\tfrac{\Delta y}{\mu}\right)^2 \tau \right).
\end{equation}

\begin{figure}
    \centering
    \includegraphics[width=1.\linewidth]{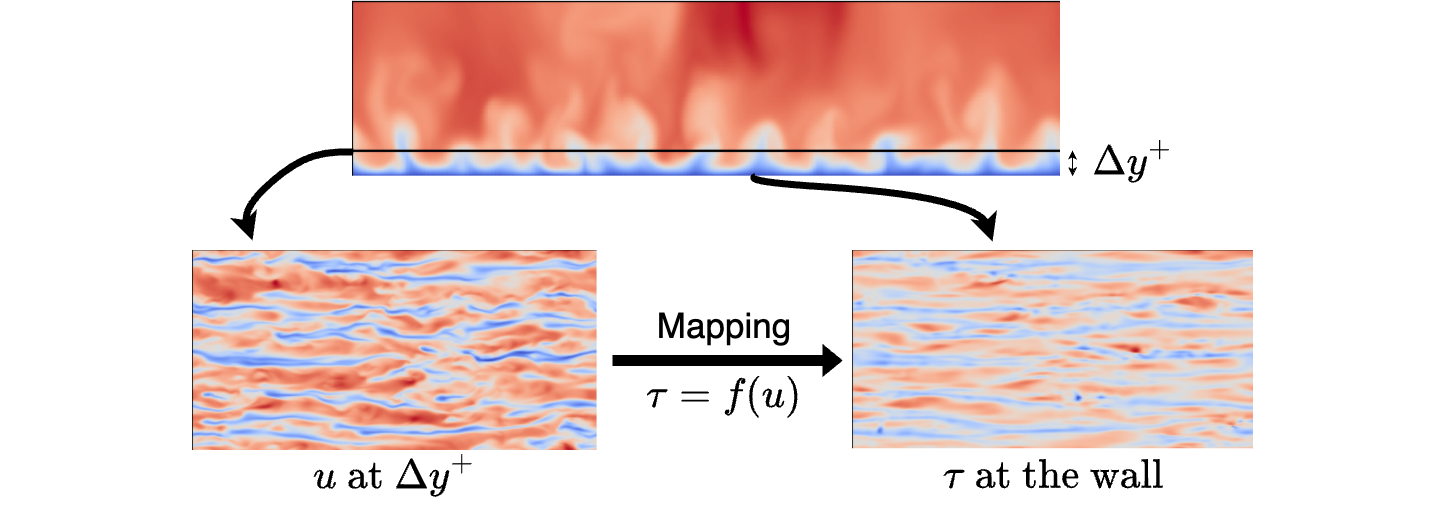}
    \caption{Representation of the wall model (side view of the channel flow at a given snapshot; top view of the velocity field $u$ off the wall and the corresponding shear stress field $\tau$)}
    \label{fig:wall_model}
\end{figure}

Figure~\ref{fig:wall_model} provides a schematic representation of the wall model, showing a side view of the instantaneous velocity field \(u\), a top view of the velocity field at a distance \(\Delta y^+\) from the wall used as input to the model, and the corresponding wall shear stress \(\tau\) to be predicted. The wall model thus represents a mapping from the velocity field to the wall shear stress. In this setting, we employ an HFNO architecture with two parallel Fourier layers: one retaining the first 8 modes, the other the subsequent 6 modes, while a linear residual predictor accounts for the remaining scales. The ranges of modes are chosen based on the spanwise correlation length (see Table~\ref{tab:metrics_models}) to isolate structures of different streak lengths observed in the flow. We also train a classical FNO model with 3 sequential Fourier layers in order to compare the obtained results.

\begin{figure}[h]
  \centering
  \includegraphics[width=0.73\textwidth]{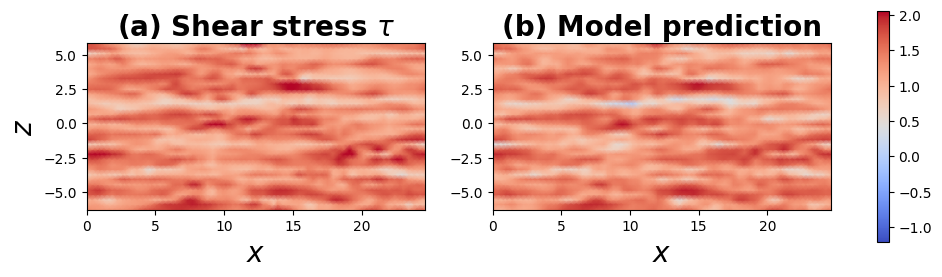}\\[1em] 
  \includegraphics[width=1\textwidth]{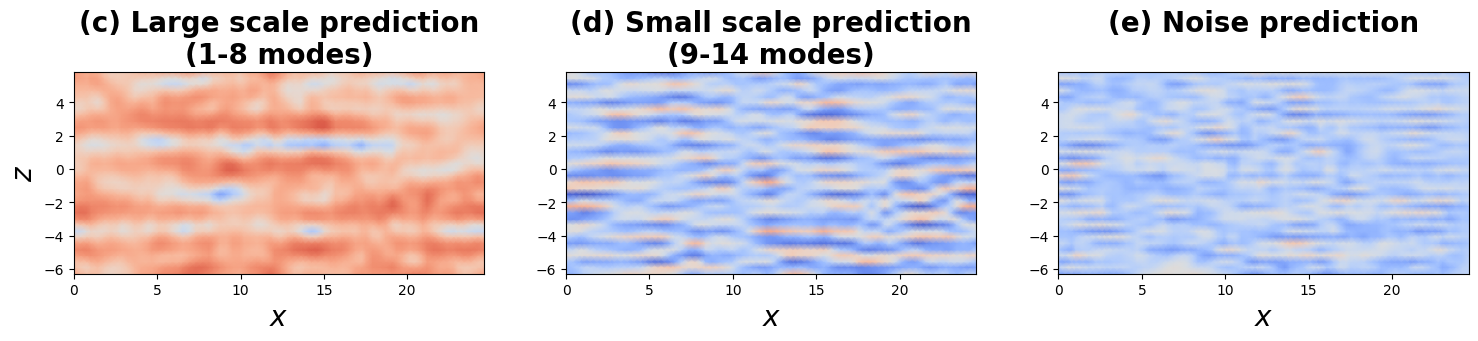}
  \caption{A random snapshot of $\tau$ at the wall, its prediction by the HFNO model and each parts of the model}
  \label{fig:wall_model_preds}
\end{figure}

\begin{figure}
    \centering
    \begin{subfigure}[b]{0.32\textwidth}
        \includegraphics[width=\textwidth]{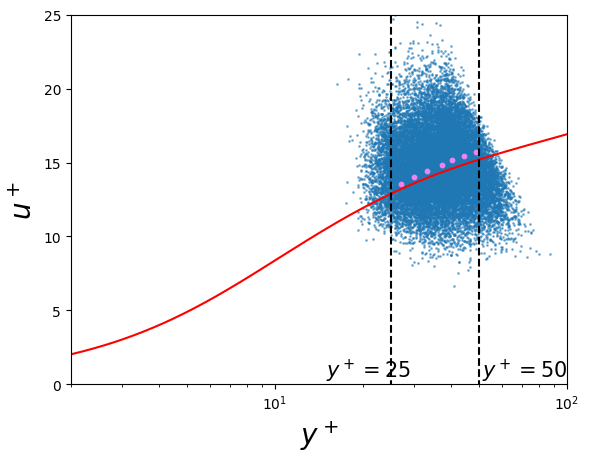}
        \caption{Ground truth}
        \label{fig:base_distrib}
    \end{subfigure}
    \begin{subfigure}[b]{0.32\textwidth}
        \includegraphics[width=\textwidth]{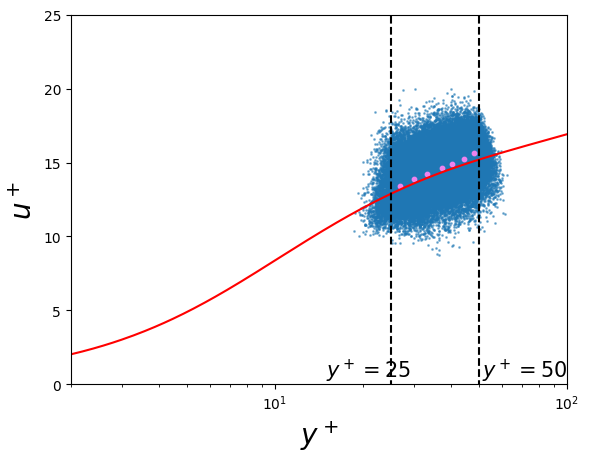}
        \caption{MLP base model}
        \label{fig:mlp_distrib}
    \end{subfigure}
    \begin{subfigure}[b]{0.32\textwidth}
        \includegraphics[width=\textwidth]{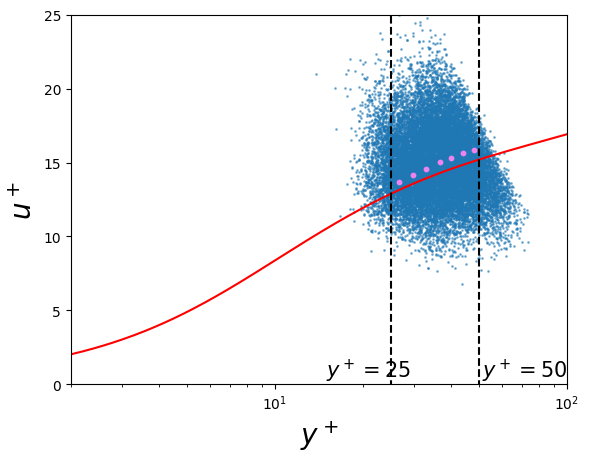}
        \caption{FNO}
        \label{fig:fno_distrib}
    \end{subfigure}

    \begin{subfigure}[b]{0.32\textwidth}
        \includegraphics[width=\textwidth]{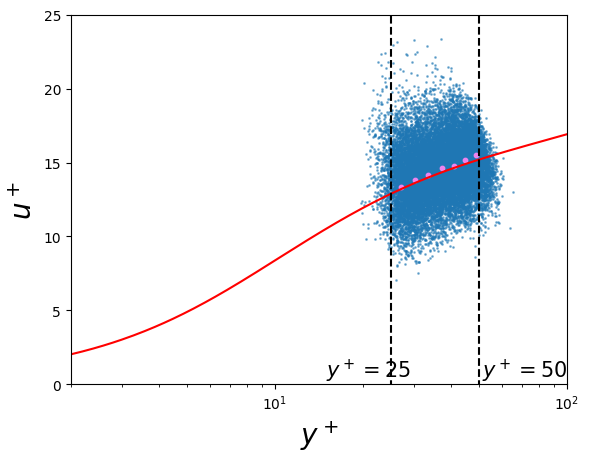}
        \caption{HFNO first layer only.}
        \label{fig:hfno1_distrib}
    \end{subfigure}
    \begin{subfigure}[b]{0.32\textwidth}
        \includegraphics[width=\textwidth]{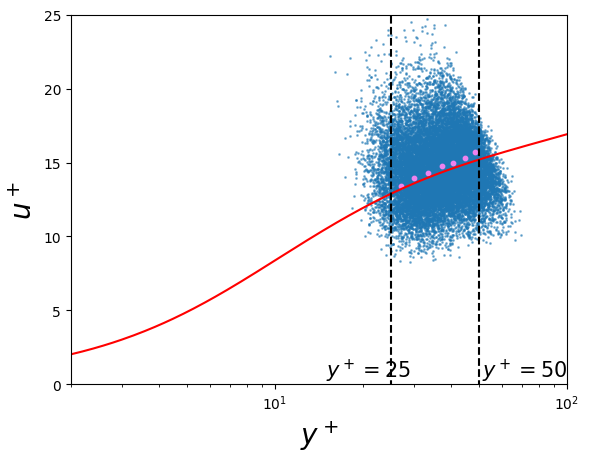}
        \caption{HFNO, two Fourier layers}
        \label{fig:hfno2_distrib}
    \end{subfigure}
    \begin{subfigure}[b]{0.32\textwidth}
        \includegraphics[width=\textwidth]{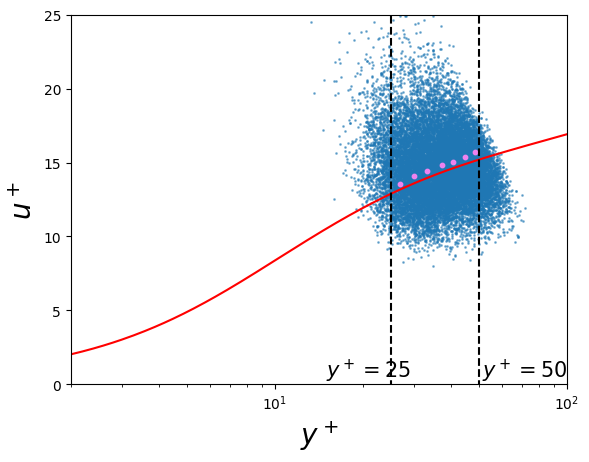}
        \caption{Complete HFNO.}
        \label{fig:hfno_distrib}
    \end{subfigure}

    \caption{$u^+(\Delta y^+)$ graphs for different predictions of $\tau$; the red curve represents the Reichardt's profile and the pink points the mean profile computed from the point cloud.}
    \label{fig:u_plus_y_plus}
\end{figure}

As illustrated in Figure~\ref{fig:wall_model_preds}, we can clearly observe that each layer captures the corresponding scales, consistent with the behavior observed in the Kuramoto--Sivashinsky and Kolmogorov flow cases.
Figure~\ref{fig:u_plus_y_plus} shows the $u^+(\Delta y^+)$ distributions obtained from different predictions of $\tau$. To generate these curves, we convert the velocity $u$ and the wall-normal distance $\Delta y$ into wall units using the corresponding shear stress $\tau$. 
The transformations are given by 
\begin{equation}
    u^+ = u \sqrt{\frac{\rho}{\tau}}, \qquad 
    \Delta y^+ = \Delta y \, \sqrt{\frac{\tau}{\rho}} \, \frac{1}{\nu}.
\end{equation}
We perform this conversion using the true $\tau$, the values predicted by the MLP baseline model, the classical FNO model, and the predictions of the HFNO model. We additionally include two intermediate
configurations of the HFNO model: one using only the first Fourier layer, and another using the first two
layers. We also compare all results to the Reichardt law \citep{reichardt_vollstandige_1951}, which has the following analytical expression:
\begin{equation}
    u^+(\Delta y^+) = \frac{1}{\kappa} \log(1 + \kappa \Delta y^+) 
    + 7.8 \left[ 1 - \exp\left(-\frac{\Delta y^+}{11}\right) 
    - \frac{\Delta y^+}{11} \exp\left(-\frac{\Delta y^+}{3}\right) \right].
\end{equation}
with $\kappa=0.41$ the Von Kármán constant. For both models, the mean profiles (shown as pink points) appear to follow the Reichardt law reasonably well, although a significant variance is observed, similar to that obtained using the true $\tau$ distribution. It is worth noting that the FNO and HFNO models exhibit a wider variance compared to the local MLP model, which allows it to capture a broader range of flow behaviors. We also note that truncated HFNO models, utilizing only the initial layers, remain competitive with the MLP baseline. While they exhibit some loss in variance prediction relative to the full HFNO or FNO models, their mean predictions are still satisfactory.

\begin{table}[htbp]
    \centering
    \caption{Comparison of five models across five metrics}
    \label{tab:metrics_models}
    \begin{tabular}{lccccc} 
        \hline
        Model & RMSE (\%) & $\rho$ & Mean error (\%) & Energy distance & Wasserstein distance \\
        \hline
        MLP base model & 25.44 & 0.56 & 0.947 & 0.129 & 2.262 $\pm$ 0.117 \\
    FNO & \textbf{5.48} & \textbf{0.98} & 1.205 & \textbf{0.009} & \textbf{0.775 $\pm$ 0.161} \\
        HFNO (1) & 24.39 & 0.58 & 1.993 & 0.099 & 1.925 $\pm$ 0.103 \\
        HFNO (1+2) & 16.71 & 0.83 & 0.947 & 0.024 & 1.086 $\pm$ 0.149 \\
        HFNO & 15.10 & 0.86 & \textbf{0.536} & 0.010 & 0.931 $\pm$ 0.112 \\
        \hline
    \end{tabular}
\end{table}

\begin{figure}
    \centering
    \begin{subfigure}[b]{0.32\textwidth}
        \includegraphics[width=\textwidth]{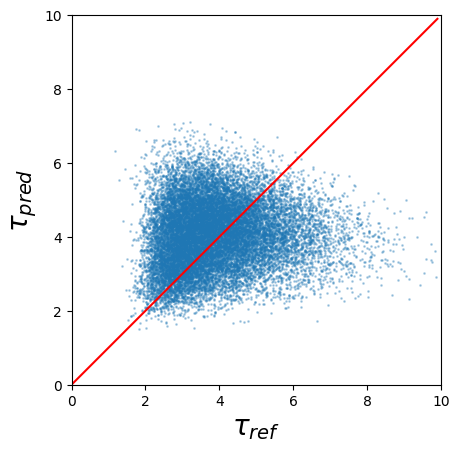}
        \caption{Reichardt's Law}
        \label{fig:RL_parity}
    \end{subfigure}
    \begin{subfigure}[b]{0.32\textwidth}
        \includegraphics[width=\textwidth]{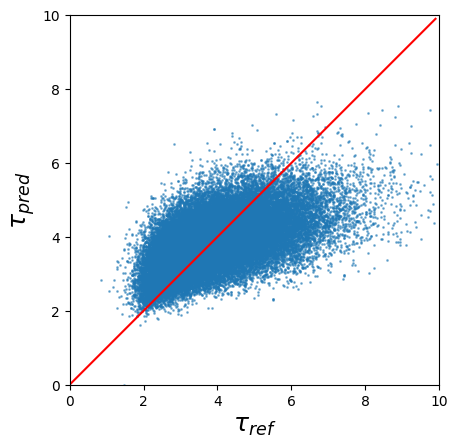}
        \caption{MLP base model}
        \label{fig:mlp_parity}
    \end{subfigure}
    \begin{subfigure}[b]{0.32\textwidth}
        \includegraphics[width=\textwidth]{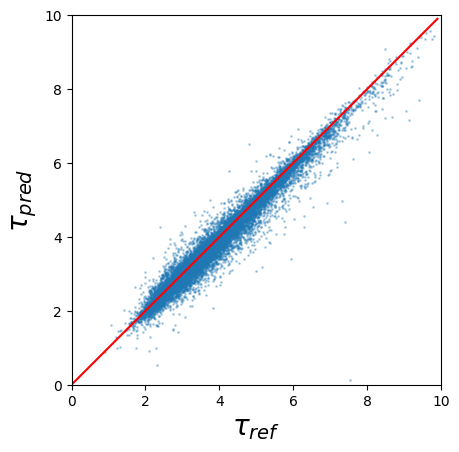}
        \caption{FNO}
        \label{fig:fno_parity}
    \end{subfigure}

    \begin{subfigure}[b]{0.32\textwidth}
        \includegraphics[width=\textwidth]{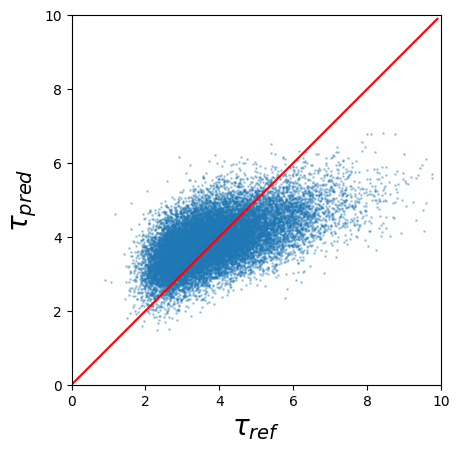}
        \caption{HFNO, first layer only.}
        \label{fig:hfno1_parity}
    \end{subfigure}
    \begin{subfigure}[b]{0.32\textwidth}
        \includegraphics[width=\textwidth]{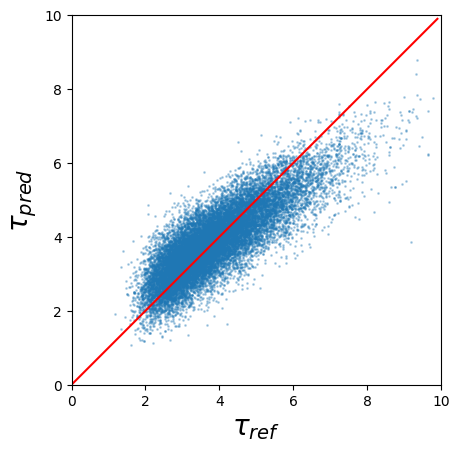}
        \caption{HFNO, two Fourier layers.}
        \label{fig:hfno2_parity}
    \end{subfigure}
    \begin{subfigure}[b]{0.32\textwidth}
        \includegraphics[width=\textwidth]{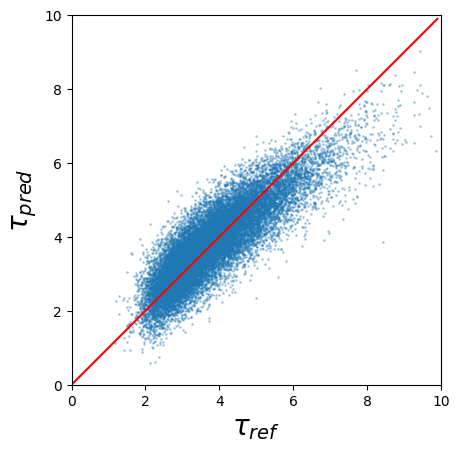}
        \caption{Complete HFNO.}
        \label{fig:hfno_parity}
    \end{subfigure}

    \caption{Parity plots for several predictions of $\tau$; the red curve represents $y=x$.}
    \label{fig:parity_plots}
\end{figure}

It is also insightful to examine the parity plot of the wall shear stress $\tau$, which compares the model predictions with the ground truth. Figure~\ref{fig:parity_plots} presents these parity plots for our model in comparison with the local MLP baseline. Each point corresponds to the predicted value of $\tau$ at a grid location for a single snapshot, and the red line represents the reference line $y = x$. The closer the point cloud lies to this line, the more accurate the model predictions are. We again include the two intermediate configurations of the HFNO model. The results clearly show how the prediction accuracy improves as additional layers and modes are incorporated. Notably, even when using only the first set of modes, our model achieves performance comparable to that of the MLP model. As expected, the FNO achieves very high accuracy. As mentioned earlier, this can be explained in part by the loss of generality caused by the decorrelation between scales in the HFNO. However, this is mainly due to the greater depth of the FNO compared to the HFNO, which remains shallow. Table~\ref{tab:metrics_models} reports several quantitative metrics that confirm the visual observations drawn from the different figures. 
For the MLP baseline, the classical FNO, and the HFNO together with its intermediate variants (HFNO1 uses only the first layer, and HFNO2 the first two layers), 
we compute five different metrics.

The first two metrics assess the model performance in terms of pointwise prediction accuracy, which corresponds to what the models are directly trained for. 
The relative Root Mean Square Error (RMSE) is defined as
\begin{equation}
\mathrm{RMSE} = 
\sqrt{
\frac{\sum_{i=1}^N \left( \tau(\mathbf{x}_i) - \hat{\tau}(\mathbf{x}_i) \right)^2}
     {\sum_{i=1}^N \tau(\mathbf{x}_i)^2}
},
\end{equation}
and the correlation coefficient $\rho$ is given by
\begin{equation}
\rho = 
\frac{
\langle \tau \hat{\tau} \rangle - \langle \tau \rangle \langle \hat{\tau} \rangle
}{
\sqrt{ \langle \tau^2 \rangle - \langle \tau \rangle^2 }
\sqrt{ \langle \hat{\tau}^2 \rangle - \langle \hat{\tau} \rangle^2 }
},
\end{equation}
where $\langle \cdot \rangle$ denotes the spatial mean over the domain.

The remaining three metrics quantify differences in the distributions of points in the $u^+(y^+)$ diagrams shown in Figure~\ref{fig:u_plus_y_plus}. 
For each model, we obtain a distribution in $\mathbb{R}^2$ that we compare with the ground truth. 
The mean error corresponds to the RMSE between the mean profiles, 
the energy distance \citep{rizzo_energy_2016} measures the discrepancy between two probability distributions $\mu$ and $\nu$, and is defined as
\begin{equation}
E(X,Y) = 2\,\mathbb{E}\|X - Y\| 
- \mathbb{E}\|X - X'\| 
- \mathbb{E}\|Y - Y'\|,
\end{equation}
where $X$ and $X'$ are i.i.d. random variables following $\mu$, and $Y$ and $Y'$ are i.i.d. random variables following $\nu$. 
Finally, the Wasserstein distance \citep{villani_optimal_2009, peyre_computational_2020} quantifies the optimal transport cost required to transform one distribution into the other.

These results indicate that, although the FNO remains the most accurate model in terms of pointwise prediction, the HFNO achieves statistically comparable performance. On average, the HFNO yields slightly better results than the FNO, which appears to exhibit a systematic bias, while both models perform similarly in terms of distributional metrics. Moreover, these results were obtained with a significantly more compact architecture: the trained FNO contains approximately 5 million parameters, compared to only 1 million for the HFNO. Interestingly, this number can be further reduced by considering truncated versions of the HFNO (HFNO1 and HFNO2 in Table~\ref{tab:metrics_models}), which remain competitive and satisfactory depending on the desired level of accuracy.

\section{Conclusion}
\label{sec:conclusion}

In this work, we have presented a Hierarchical Fourier Neural Operator (HFNO) architecture tailored to dynamical systems with multiscale spatial structures. By decomposing the Fourier modes into separate layers corresponding to physically meaningful ranges — energy-containing, inertial, and dissipative scales — the model is able to capture the dominant dynamics at each scale. The use of MLPs for each Fourier layer, complemented by a residual predictor that can take different forms, allows the network to faithfully reconstruct both large-scale coherent structures and small-scale fluctuations.

Applications to the Kuramoto–Sivashinsky equation, the Kolmogorov flow, and wall modelling in a turbulent channel flow demonstrate the effectiveness of the proposed approach: each Fourier layer predominantly contributes to its assigned wavenumber range, leading to accurate predictions and
generalization capabilities. The hierarchical design not only improves interpretability by linking model components to physical scales, but also enhances efficiency by reducing the complexity of each sub-task.

Overall, these results highlight the potential of HFNOs, for modeling complex turbulent flows in a meshless and physics-informed manner. Future work will focus on extending this framework to higher Reynolds numbers and more realistic three-dimensional flows, as well as exploring adaptive strategies for mode separation to further optimize performance across a broader range of dynamical regimes.

\begin{Backmatter}

\paragraph{Acknowledgments} The authors gratefully acknowledge Ardeshir Hanifi, Ricardo Vinuesa, and Arivazhagan Geetha Balasubramanian from the Division of Fluid Mechanics, KTH Royal Institute of Technology (Sweden), for kindly providing the open-channel dataset used in this study.

\paragraph{Funding Statement}
This research was supported by grant from the Agence Nationale de la Recherche (ANR), under the funding ID:
ANR-22-CE94-0001; EDNNForest.

\paragraph{Competing Interests}
None

\paragraph{Data Availability Statement}
Replication data and code can be found: \url{https://github.com/marcocayuela/HFNO}

\paragraph{Ethical Standards}
The research meets all ethical guidelines, including adherence to the legal requirements of the study country.

\paragraph{Author Contributions}
\textbf{Marco Cayuela } : Conceptualization, Methodology, Data curation, Data Visualisation, Formal analysis, Validation, Writing original draft. \textbf{Vincent Le Chenadec} : Resources, Supervision, Validation, Writing – review \& editing.  \textbf{Peter J. Schmid} : Resources, Supervision, Validation, Writing – review \& editing. \textbf{Taraneh Sayadi} :  Conceptualization, Methodology, Resources, Formal analysis, Supervision, Validation, Writing – review \& editing. \\
All authors approved the final submitted draft.


\printbibliography

@misc{li_fourier_2021,
	title = {Fourier {Neural} {Operator} for {Parametric} {Partial} {Differential} {Equations}},
	url = {http://arxiv.org/abs/2010.08895},
	urldate = {2023-12-05},
	publisher = {arXiv},
	author = {Li, Zongyi and Kovachki, Nikola and Azizzadenesheli, Kamyar and Liu, Burigede and Bhattacharya, Kaushik and Stuart, Andrew and Anandkumar, Anima},
	month = may,
	year = {2021},
	note = {arXiv:2010.08895 [cs, math]},
}

@article{bae_dynamic_2019,
	title = {Dynamic slip wall model for large-eddy simulation},
	volume = {859},
	issn = {0022-1120, 1469-7645},
	url = {http://arxiv.org/abs/1810.04214},
	doi = {10.1017/jfm.2018.838},
	urldate = {2024-05-06},
	journal = {Journal of Fluid Mechanics},
	author = {Bae, H. Jane and Lozano-Durán, Adrián and Bose, Sanjeeb T. and Moin, Parviz},
	month = jan,
	year = {2019},
	note = {arXiv:1810.04214 [physics]},
	pages = {400--432},
}

@article{dupuy_data-driven_2023,
	title = {Data-driven wall modeling for turbulent separated flows},
	volume = {487},
	issn = {00219991},
	url = {https://linkinghub.elsevier.com/retrieve/pii/S0021999123002681},
	doi = {10.1016/j.jcp.2023.112173},
	language = {en},
	urldate = {2024-06-24},
	journal = {Journal of Computational Physics},
	author = {Dupuy, D. and Odier, N. and Lapeyre, C.},
	month = aug,
	year = {2023},
	pages = {112173},
}

@article{choi_grid-point_2012,
	title = {Grid-point requirements for large eddy simulation: {Chapman}’s estimates revisited},
	volume = {24},
	issn = {1070-6631, 1089-7666},
	shorttitle = {Grid-point requirements for large eddy simulation},
	url = {https://pubs.aip.org/pof/article/24/1/011702/361102/Grid-point-requirements-for-large-eddy-simulation},
	doi = {10.1063/1.3676783},
	language = {en},
	number = {1},
	urldate = {2024-06-25},
	journal = {Physics of Fluids},
	author = {Choi, Haecheon and Moin, Parviz},
	month = jan,
	year = {2012},
	pages = {011702},
}

@article{moin_direct_1998,
	title = {{DIRECT} {NUMERICAL} {SIMULATION}: {A} {Tool} in {Turbulence} {Research}},
	volume = {30},
	issn = {0066-4189, 1545-4479},
	shorttitle = {{DIRECT} {NUMERICAL} {SIMULATION}},
	url = {https://www.annualreviews.org/content/journals/10.1146/annurev.fluid.30.1.539},
	doi = {10.1146/annurev.fluid.30.1.539},
	language = {fr},
	number = {Volume 30, 1998},
	urldate = {2025-07-03},
	journal = {Annual Review of Fluid Mechanics},
	author = {Moin, Parviz and {Krishnan Mahesh}},
	month = jan,
	year = {1998},
	note = {Publisher: Annual Reviews},
	pages = {539--578},
}

@misc{pope_turbulent_2000,
	title = {Turbulent {Flows}},
	url = {https://www.cambridge.org/highereducation/books/turbulent-flows/C58EFF59AF9B81AE6CFAC9ED16486B3A},
	language = {en},
	urldate = {2025-07-03},
	journal = {Higher Education from Cambridge University Press},
	author = {Pope, Stephen B.},
	month = aug,
	year = {2000},
	doi = {10.1017/CBO9780511840531},
	note = {ISBN: 9780511840531
Publisher: Cambridge University Press},
}

@article{lecun_gradient-based_1998,
	title = {Gradient-based learning applied to document recognition},
	volume = {86},
	copyright = {https://ieeexplore.ieee.org/Xplorehelp/downloads/license-information/IEEE.html},
	issn = {00189219},
	url = {http://ieeexplore.ieee.org/document/726791/},
	doi = {10.1109/5.726791},
	number = {11},
	urldate = {2025-07-03},
	journal = {Proceedings of the IEEE},
	author = {Lecun, Y. and Bottou, L. and Bengio, Y. and Haffner, P.},
	month = nov,
	year = {1998},
	pages = {2278--2324},
}

@article{jaeger_echo_2010,
	title = {The “echo state” approach to analysing and training recurrent neural networks – with an {Erratum} note},
	language = {en},
	author = {Jaeger, Herbert},
	year = {2010},
}

@book{canuto_spectral_1988,
	address = {Berlin, Heidelberg},
	title = {Spectral {Methods} in {Fluid} {Dynamics}},
	copyright = {http://www.springer.com/tdm},
	isbn = {978-3-540-52205-8 978-3-642-84108-8},
	url = {http://link.springer.com/10.1007/978-3-642-84108-8},
	urldate = {2025-07-23},
	publisher = {Springer},
	author = {Canuto, Claudio and Hussaini, M. Yousuff and Quarteroni, Alfio and Zang, Thomas A.},
	year = {1988},
	doi = {10.1007/978-3-642-84108-8},
}

@misc{magri_luca_github_2021,
	title = {{GitHub} - {MagriLab}/{KolSol}: {Pseudospectral} {Kolmogorov} {Flow} {Solver}},
	url = {https://github.com/MagriLab/KolSol?tab=readme-ov-file#R1},
	urldate = {2025-07-23},
	author = {{Magri Luca}},
	year = {2021},
}

@article{lu_deeponet_2021,
	title = {{DeepONet}: {Learning} nonlinear operators for identifying differential equations based on the universal approximation theorem of operators},
	volume = {3},
	issn = {2522-5839},
	shorttitle = {{DeepONet}},
	url = {http://arxiv.org/abs/1910.03193},
	doi = {10.1038/s42256-021-00302-5},
	number = {3},
	urldate = {2025-10-07},
	journal = {Nature Machine Intelligence},
	author = {Lu, Lu and Jin, Pengzhan and Karniadakis, George Em},
	month = mar,
	year = {2021},
	note = {arXiv:1910.03193 [cs]},
	pages = {218--229},
}

@article{raissi_physics-informed_2019,
	title = {Physics-informed neural networks: {A} deep learning framework for solving forward and inverse problems involving nonlinear partial differential equations},
	volume = {378},
	issn = {0021-9991},
	shorttitle = {Physics-informed neural networks},
	url = {https://www.sciencedirect.com/science/article/pii/S0021999118307125},
	doi = {10.1016/j.jcp.2018.10.045},
	urldate = {2025-10-08},
	journal = {Journal of Computational Physics},
	author = {Raissi, M. and Perdikaris, P. and Karniadakis, G. E.},
	month = feb,
	year = {2019},
	pages = {686--707},
}

@article{piomelli_large-eddy_1999,
	title = {Large-eddy simulation: achievements and challenges},
	volume = {35},
	issn = {0376-0421},
	shorttitle = {Large-eddy simulation},
	url = {https://www.sciencedirect.com/science/article/pii/S0376042198000141},
	doi = {10.1016/S0376-0421(98)00014-1},
	number = {4},
	urldate = {2025-10-20},
	journal = {Progress in Aerospace Sciences},
	author = {Piomelli, U.},
	month = may,
	year = {1999},
	pages = {335--362},
}

@book{ian_goodfellow_deep_2016,
	edition = {MIT Press},
	title = {Deep {Learning}},
	url = {http://www.deeplearningbook.org},
	author = {{Ian Goodfellow} and {Yoshua Bengio} and {Aaron Courville}},
	year = {2016},
}

@misc{kingma_auto-encoding_2022,
	title = {Auto-{Encoding} {Variational} {Bayes}},
	url = {http://arxiv.org/abs/1312.6114},
	doi = {10.48550/arXiv.1312.6114},
	urldate = {2025-10-20},
	publisher = {arXiv},
	author = {Kingma, Diederik P. and Welling, Max},
	year = {2022},
	note = {arXiv:1312.6114 [stat]},
}

@article{hinton_reducing_2006,
	title = {Reducing the dimensionality of data with neural networks},
	volume = {313},
	issn = {1095-9203},
	doi = {10.1126/science.1127647},
	language = {eng},
	number = {5786},
	journal = {Science (New York, N.Y.)},
	author = {Hinton, G. E. and Salakhutdinov, R. R.},
	month = jul,
	year = {2006},
	pmid = {16873662},
	pages = {504--507},
}

@article{ozalp_stability_2025,
	title = {Stability analysis of chaotic systems in latent spaces},
	volume = {113},
	issn = {1573-269X},
	url = {https://doi.org/10.1007/s11071-024-10712-w},
	doi = {10.1007/s11071-024-10712-w},
	language = {en},
	number = {11},
	urldate = {2025-10-20},
	journal = {Nonlinear Dynamics},
	author = {Özalp, Elise and Magri, Luca},
	month = jun,
	year = {2025},
	pages = {13791--13806},
}

@misc{pathak_fourcastnet_2022,
	title = {{FourCastNet}: {A} {Global} {Data}-driven {High}-resolution {Weather} {Model} using {Adaptive} {Fourier} {Neural} {Operators}},
	shorttitle = {{FourCastNet}},
	url = {http://arxiv.org/abs/2202.11214},
	doi = {10.48550/arXiv.2202.11214},
	urldate = {2025-10-22},
	publisher = {arXiv},
	author = {Pathak, Jaideep and Subramanian, Shashank and Harrington, Peter and Raja, Sanjeev and Chattopadhyay, Ashesh and Mardani, Morteza and Kurth, Thorsten and Hall, David and Li, Zongyi and Azizzadenesheli, Kamyar and Hassanzadeh, Pedram and Kashinath, Karthik and Anandkumar, Animashree},
	month = feb,
	year = {2022},
	note = {arXiv:2202.11214 [physics]},
}

@misc{kovachki_neural_2024,
	title = {Neural {Operator}: {Learning} {Maps} {Between} {Function} {Spaces}},
	shorttitle = {Neural {Operator}},
	url = {http://arxiv.org/abs/2108.08481},
	doi = {10.5555/3648699.3648788},
	urldate = {2025-10-22},
	author = {Kovachki, Nikola and Li, Zongyi and Liu, Burigede and Azizzadenesheli, Kamyar and Bhattacharya, Kaushik and Stuart, Andrew and Anandkumar, Anima},
	month = may,
	year = {2024},
	note = {arXiv:2108.08481 [cs]},
}

@article{heyder_echo_2021,
	title = {Echo {State} {Network} for two-dimensional turbulent moist {Rayleigh}-{Bénard} convection},
	volume = {103},
	issn = {2470-0045, 2470-0053},
	url = {http://arxiv.org/abs/2101.11325},
	doi = {10.1103/PhysRevE.103.053107},
	language = {en},
	number = {5},
	urldate = {2025-10-22},
	journal = {Physical Review E},
	author = {Heyder, Florian and Schumacher, Jörg},
	month = may,
	year = {2021},
	note = {arXiv:2101.11325 [physics]},
	pages = {053107},
}

@article{reichardt_vollstandige_1951,
	title = {Vollständige {Darstellung} der turbulenten {Geschwindigkeitsverteilung} in glatten {Leitungen}},
	volume = {31},
	copyright = {Copyright © 1951 WILEY-VCH Verlag GmbH \& Co. KGaA, Weinheim},
	issn = {1521-4001},
	url = {https://onlinelibrary.wiley.com/doi/abs/10.1002/zamm.19510310704},
	doi = {10.1002/zamm.19510310704},
	language = {de},
	number = {7},
	urldate = {2025-10-23},
	journal = {ZAMM - Journal of Applied Mathematics and Mechanics / Zeitschrift für Angewandte Mathematik und Mechanik},
	author = {Reichardt, H.},
	year = {1951},
	note = {\_eprint: https://onlinelibrary.wiley.com/doi/pdf/10.1002/zamm.19510310704},
	pages = {208--219},
}

@article{choubineh_fourier_2023,
	title = {Fourier {Neural} {Operator} for {Fluid} {Flow} in {Small}-{Shape} {2D} {Simulated} {Porous} {Media} {Dataset}},
	volume = {16},
	copyright = {http://creativecommons.org/licenses/by/3.0/},
	issn = {1999-4893},
	url = {https://www.mdpi.com/1999-4893/16/1/24},
	doi = {10.3390/a16010024},
	language = {en},
	number = {1},
	urldate = {2025-10-23},
	journal = {Algorithms},
	author = {Choubineh, Abouzar and Chen, Jie and Wood, David A. and Coenen, Frans and Ma, Fei},
	month = jan,
	year = {2023},
	note = {Publisher: Multidisciplinary Digital Publishing Institute},
	pages = {24},
}

@article{wang_prediction_2024,
	title = {Prediction of turbulent channel flow using {Fourier} neural operator-based machine-learning strategy},
	volume = {9},
	doi = {10.1103/PhysRevFluids.9.084604},
	number = {8},
	journal = {Physical Review Fluids},
	author = {Wang, Yunpeng},
	year = {2024},
}

@misc{serrano_infinity_2023,
	title = {{INFINITY}: {Neural} {Field} {Modeling} for {Reynolds}-{Averaged} {Navier}-{Stokes} {Equations}},
	shorttitle = {{INFINITY}},
	url = {http://arxiv.org/abs/2307.13538},
	doi = {10.48550/arXiv.2307.13538},
	urldate = {2025-10-24},
	publisher = {arXiv},
	author = {Serrano, Louis and Migus, Leon and Yin, Yuan and Mazari, Jocelyn Ahmed and Gallinari, Patrick},
	month = jul,
	year = {2023},
	note = {arXiv:2307.13538 [cs]},
}

@article{launder_numerical_1974,
	title = {The numerical computation of turbulent flows},
	volume = {3},
	issn = {0045-7825},
	url = {https://www.sciencedirect.com/science/article/pii/0045782574900292},
	doi = {10.1016/0045-7825(74)90029-2},
	number = {2},
	urldate = {2025-10-24},
	journal = {Computer Methods in Applied Mechanics and Engineering},
	author = {Launder, B. E. and Spalding, D. B.},
	month = mar,
	year = {1974},
	pages = {269--289},
}

@misc{li_neural_2020,
	title = {Neural {Operator}: {Graph} {Kernel} {Network} for {Partial} {Differential} {Equations}},
	shorttitle = {Neural {Operator}},
	url = {http://arxiv.org/abs/2003.03485},
	doi = {10.48550/arXiv.2003.03485},
	urldate = {2025-10-24},
	publisher = {arXiv},
	author = {Li, Zongyi and Kovachki, Nikola and Azizzadenesheli, Kamyar and Liu, Burigede and Bhattacharya, Kaushik and Stuart, Andrew and Anandkumar, Anima},
	month = mar,
	year = {2020},
	note = {arXiv:2003.03485 [cs]},
}

@book{sagaut_pierre_large_2006,
	address = {Berlin/Heidelberg},
	series = {Scientific {Computation}},
	title = {Large {Eddy} {Simulation} for {Incompressible} {Flows}},
	copyright = {http://www.springer.com/tdm},
	isbn = {978-3-540-26344-9},
	url = {http://link.springer.com/10.1007/b137536},
	language = {en},
	urldate = {2025-10-24},
	publisher = {Springer-Verlag},
	author = {{Sagaut Pierre}},
	year = {2006},
	doi = {10.1007/b137536},
}

@article{meneveau_scale-invariance_2000,
	title = {Scale-{Invariance} and {Turbulence} {Models} for {Large}-{Eddy} {Simulation}},
	volume = {32},
	issn = {0066-4189, 1545-4479},
	url = {https://www.annualreviews.org/content/journals/10.1146/annurev.fluid.32.1.1},
	doi = {10.1146/annurev.fluid.32.1.1},
	language = {en},
	number = {Volume 32, 2000},
	urldate = {2025-10-24},
	journal = {Annual Review of Fluid Mechanics},
	author = {Meneveau, Charles and Katz, Joseph},
	month = jan,
	year = {2000},
	note = {Publisher: Annual Reviews},
	pages = {1--32},
}

@article{tripura_wavelet_2023,
	title = {Wavelet {Neural} {Operator} for solving parametric partial differential equations in computational mechanics problems},
	volume = {404},
	issn = {0045-7825},
	url = {https://www.sciencedirect.com/science/article/pii/S0045782522007393},
	doi = {10.1016/j.cma.2022.115783},
	urldate = {2025-10-24},
	journal = {Computer Methods in Applied Mechanics and Engineering},
	author = {Tripura, Tapas and Chakraborty, Souvik},
	month = feb,
	year = {2023},
	pages = {115783},
}

@article{hornik_multilayer_1989,
	title = {Multilayer feedforward networks are universal approximators},
	volume = {2},
	issn = {0893-6080},
	url = {https://www.sciencedirect.com/science/article/pii/0893608089900208},
	doi = {10.1016/0893-6080(89)90020-8},
	number = {5},
	urldate = {2025-10-24},
	journal = {Neural Networks},
	author = {Hornik, Kurt and Stinchcombe, Maxwell and White, Halbert},
	month = jan,
	year = {1989},
	pages = {359--366},
}

@book{chevalier_mattias_simson_2007,
	address = {Stockholm},
	title = {{SIMSON}: a pseudo-spectral solver for incompressible boundary layer flows},
	isbn = {978-91-7178-838-2},
	shorttitle = {{SIMSON}},
	language = {en},
	publisher = {Mekanik, Kungliga Tekniska högskolan},
	author = {{Chevalier Mattias}},
	year = {2007},
}

@article{cabot_approximate_2000,
	title = {Approximate {Wall} {Boundary} {Conditions} in the {Large}-{Eddy} {Simulation} of {High} {Reynolds} {Number} {Flow}},
	volume = {63},
	issn = {1573-1987},
	url = {https://doi.org/10.1023/A:1009958917113},
	doi = {10.1023/A:1009958917113},
	language = {en},
	number = {1},
	urldate = {2025-10-28},
	journal = {Flow, Turbulence and Combustion},
	author = {Cabot, W. and Moin, P.},
	month = jan,
	year = {2000},
	pages = {269--291},
}

@article{rizzo_energy_2016,
	title = {Energy distance},
	volume = {8},
	copyright = {http://onlinelibrary.wiley.com/termsAndConditions\#vor},
	issn = {1939-5108, 1939-0068},
	url = {https://wires.onlinelibrary.wiley.com/doi/10.1002/wics.1375},
	doi = {10.1002/wics.1375},
	language = {en},
	number = {1},
	urldate = {2025-10-31},
	journal = {WIREs Computational Statistics},
	author = {Rizzo, Maria L. and Székely, Gábor J.},
	month = jan,
	year = {2016},
	pages = {27--38},
}

@book{villani_optimal_2009,
	address = {Berlin, Heidelberg},
	series = {Grundlehren der mathematischen {Wissenschaften}},
	title = {Optimal {Transport}},
	volume = {338},
	copyright = {http://www.springer.com/tdm},
	isbn = {978-3-540-71049-3 978-3-540-71050-9},
	url = {http://link.springer.com/10.1007/978-3-540-71050-9},
	urldate = {2025-10-31},
	publisher = {Springer},
	author = {Villani, Cédric},
	editor = {Berger, M. and Eckmann, B. and De La Harpe, P. and Hirzebruch, F. and Hitchin, N. and Hörmander, L. and Kupiainen, A. and Lebeau, G. and Ratner, M. and Serre, D. and Sinai, Ya. G. and Sloane, N. J. A. and Vershik, A. M. and Waldschmidt, M.},
	year = {2009},
	doi = {10.1007/978-3-540-71050-9},
}

@misc{peyre_computational_2020,
	title = {Computational {Optimal} {Transport}},
	url = {http://arxiv.org/abs/1803.00567},
	doi = {10.48550/arXiv.1803.00567},
	urldate = {2025-10-31},
	publisher = {arXiv},
	author = {Peyré, Gabriel and Cuturi, Marco},
	month = mar,
	year = {2020},
	note = {arXiv:1803.00567 [stat]},
}

\end{Backmatter}

\end{document}